\documentclass[letterpaper]{article} 
\usepackage[draft]{aaai25}  
\usepackage{times}  
\usepackage{helvet}  
\usepackage{courier}  
\usepackage[hyphens]{url}  
\usepackage{graphicx} 
\usepackage{natbib}  
\usepackage{caption} 
\usepackage{algorithm}
\usepackage{algorithmic}

\usepackage{newfloat}
\usepackage{listings}
\DeclareCaptionStyle{ruled}{labelfont=normalfont,labelsep=colon,strut=off} 
\floatstyle{ruled}
\newfloat{listing}{tb}{lst}{}
\floatname{listing}{Listing}
\usepackage{multirow,array, makecell}
\usepackage{amsthm,amsmath, xcolor, appendix}
\usepackage{amsfonts, tikz}
\usetikzlibrary{positioning, calc}
\usetikzlibrary {decorations.pathreplacing}
\usepackage[font=small]{caption}
\usepackage{subcaption}

\DeclareMathOperator*{\argmax}{arg\,max}

\newtheorem{innercustomex}{Example}
\newtheorem{innercustomprp}{Proposition}
\newtheorem{innercustomlem}{Lemma}
\newtheorem{innercustomthrm}{Theorem}

\newenvironment{customthrm}[1]
  {\renewcommand\theinnercustomthrm{#1}\innercustomthrm}
  {\endinnercustomthrm}

\newenvironment{customprp}[1]
  {\renewcommand\theinnercustomprp{#1}\innercustomprp}
  {\endinnercustomprp}
\newenvironment{customex}[1]
  {\renewcommand\theinnercustomex{#1}\innercustomex}
  {\endinnercustomex}
  \newenvironment{customlem}[1]
  {\renewcommand\theinnercustomlem{#1}\innercustomlem}
  {\endinnercustomlem}

\title{Equal Merit Does Not Imply Equality: Discrimination at Equilibrium\\ in a Hiring Market with Symmetric Agents}

\author {
    Serafina Kamp\textsuperscript{\rm 1},
    Benjamin Fish\textsuperscript{\rm 1}
}
\affiliations {
    \textsuperscript{\rm 1}University of Michigan Department of Computer Science Engineering\\
    serafibk@umich.edu, benfish@umich.edu
}

\begin{document}

\maketitle

\begin{abstract}
Machine learning has grown in popularity to help assign resources and make decisions about users, which can result in discrimination. This includes hiring markets, where employers have increasingly been interested in using automated tools to help hire candidates. In response, there has been significant effort to understand and mitigate the sources of discrimination in these tools.  However, previous work has largely assumed that discrimination, in any area of ML, is the result of some initial \textit{unequal distribution of resources} across groups: One group is on average less qualified, there is less training data for one group, or the classifier is less accurate on one group, etc. However, recent work have suggested that there are other sources of discrimination, such as relational inequality, that are notably non-distributional. First, we show consensus in strategy choice is a non-distributional source of inequality at equilibrium in games: We provide subgame perfect equilibria in a simple sequential model of a hiring market with Rubinstein-style bargaining between firms and candidates that exhibits asymmetric wages resulting from differences in agents' threat strategies during bargaining. 
Second, we give an initial analysis of how agents could learn such strategies via convergence of an online learning algorithm to asymmetric equilibria. Ultimately, this work motivates the further study of endogenous, possibly non-distributional, mechanisms of inequality in ML.
\end{abstract}

%

\section{Introduction}
Machine learning (ML) algorithms are marketed to make more efficient and data-driven decisions that improve human decision-making. These automated tools have become increasingly popular in recommendation systems, classification problems, and resource assignment amongst other areas \citep{berk2021fairness,calders2010three,kamiran2012data}. With their growing popularity, it has become clear that these algorithms can engender discrimination against individuals or demographic groups \citep{angwin2016machine,eubanks2018automating,miller2015algorithms,o2017weapons}. 

In response, there is a growing research area on formalizing, measuring, and mitigating discrimination~\citep{berk2021fairness,chouldechova2018frontiers,friedler2019comparative,mehrabi2021survey}. Here, discrimination is typically taken to mean the \textit{unequal distribution of resources among different groups} defined by sensitive attributes, such as race or gender~\citep{arrow2015theory,barocas2017fairness,hardt2016equality}. Many fairness metrics have been developed across ML and economics to measure such resource differences. They cover group-level, individual-level, and causal measurements alongside notions of taste-based and statistical discrimination \citep{arrow2015theory,dwork2012fairness,hardt2016equality,kamiran2012data,kusner2017counterfactual,phelps1972statistical}. Some examples of resources that could be distributed unequally include merit (i.e., relevant features or skills), data sets (i.e., distribution of training or ground truth labels, costs in acquiring data, or elasticity in demand for data), and opportunities (i.e., job offers, loan approvals, housing bids, or generally any ML prediction). In general, a fairness metric is satisfied if the corresponding resources it measures are approximately equal among groups or individuals of interest, a view of equality commonly referred to as distributive equality~\citep{dworkin2002sovereign}.  If there is no such inequality in resources -- everyone to be classified has equal merit -- then under this view we should expect no discrimination.

However, sources of discrimination in general, let alone in ML, need not be distributional~\citep{abbasi2019fairness,birhane2021algorithmic,elford2017survey,green2022escaping,hoffmann2019fairness,fish2022s,kasirzadeh2022algorithmic,schemmel2012distributive}. Various other kinds of inequality, including relational~\citep{anderson1999point}, representational~\citep{shelby2023sociotechnical,west2019discriminating}, status~\citep{ridgeway2014status}, and power~\citep{kasy2021fairness} have been proposed. Many of these are very difficult to measure, like relational equality, or still require existing distributional inequality, like when representational inequalities are caused by disparities in training data that an algorithm reproduces (e.g.\ searches for `CEO' in image search engines \citep{kay2015unequal,lam2018gender}).

Given the recent work that recognizes the limitations of distributive equality~\citep{birhane2021algorithmic,fish2022s, kasirzadeh2022algorithmic,kasy2021fairness}, in this paper, we provide a non-distributional source of inequality that can be measured via a game-theoretic model: the strategies of the agents at equilibrium. 


Throughout this paper, we use the setting of a hiring market where agents are a priori resource-symmetric: we assume no exogenous differences between agents in our market, including merit. Hiring markets are areas where ML is increasingly popular~\citep{bogen2018help, raghavan2020mitigating}, yet it has become clear that ML can perpetuate or engender discrimination here ~\citep{cavounidis2015discrimination,dastin2018amazon,hu2017fairness}. Further, in hiring markets, economic prospects are tied to both merit (i.e., having sufficient skills required to perform a job or being otherwise deemed by other agents to be deserving of a job) and how agents interact with others in the market. So, by assuming a priori resource-symmetric agents, we can isolate agent interactions via strategy choice as the only source of inequality at equilibrium.


We provide a model of a hiring market as a bargaining game to show the existence of asymmetric wages at equilibrium among agents with equal merit. Notably, by finding outcomes at equilibrium, the inequality will persist as the best-response for each agent in our market. Our bargaining game captures the wage negotiation process, an important process where agent interactions determine economic outcomes. We are interested in identifying a \textit{source} of inequality at equilibrium in hiring markets, and wage negotiation must happen in any hiring context unless the firm and candidate are \textit{a priori unequal} and the candidate is forced to accept any wage. We consider a two-sided market where firms are looking to fill a job position and candidates are applying for these positions. Once a firm and candidate are matched (i.e.\ the candidate applies for the firm's position), they participate in wage negotiation via a Rubinstein-style bargaining game~\citep{osborne1990bargaining} over the surplus generated by the candidate's employment. 

The asymmetric outcomes we find at equilibrium are driven by consensus among agents about their choice of strategy. Consensus over choice of strategy can be non-distributional when the consensus is not driven by exogenous disparities in merit or other such factors.  As an example, consider social norms in negotiation: Women may follow social norms where they do not negotiate for higher salaries, as compared to men, which contributes to observed pay gaps between these genders~\citep{rengender}. Here, the social norms may act as a non-distributional source of the observed discriminatory pay gaps and, further, adherence to discriminatory social norms may cause or exacerbate a gender social hierarchy~\citep{ferrant2014unpaid}. Another example could be in a consensus among firms to use an algorithm for price setting~\citep{weber2023hub}. If enough firms believe that the algorithm offers the true market value of their product, say, rental housing, then firms will all list the price given by the algorithm for their units and it will become the market value.

Therefore, it is crucial to study the conditions under which strategy consensus drives inequality at equilibrium, including via the choice to use algorithmic decisions. This paper addresses this by (1) demonstrating endogenous consensus about outside options among equal merit agents which creates wage inequality at equilibrium in a bargaining game and (2) showing initial results for how an online learning algorithm would converge to such strategies.

Section~\ref{sec:discriminatory_equilibria} gives our first result. We create a resource-symmetric market with the following assumptions 
:1) every pair of agents split equal surplus they are equally entitled to, ensuring equal merit; 2) every agent has an endogenous outside option so that every agent has an a priori chance to have bargaining power; 3) the game itself is symmetric so that no agent has a priori more bargaining power than any other, e.g.\ by getting to propose an offer first; 4) all agents are identical up to type: The only difference between agents are their strategies. We show that differences in credible threats can be a mechanism for non-distributional discrimination by showing that there exist multiple strategy profiles in subgame perfect equilibrium (SPE) in the bargaining game with asymmetric payoffs. A set of strategies is in SPE when there are no beneficial deviations from the strategies on or off the equilibrium path.\footnote{See for example~\citet{tadelis} for an introduction to the game theoretic concepts used in this work.} As such, all threats used by agents must be credible, i.e., it must be rational for them to take any action specified in their strategy. We highlight two kinds of asymmetric outcomes at equilibrium: One where a group of candidates gets more of the surplus than another group of candidates and another where the firms get more of the surplus than the candidates. 

Section~\ref{sec:learning_in_bargaining} gives our second result. Our interest here is in understanding when it is possible for online learning algorithms to converge to equilibrium strategies in a bargaining game. In this paper, we initiate a study of how learning algorithms can learn the kinds of strategies that constitute non-distributional inequality. As is typical in learning in games, we model strategy learning in bargaining games as an online optimization problem. We use a simplified version of the above bargaining game to one that has only two players and finitely many rounds.  This approach does not require access to specific wage setting algorithms used in practice, which are often proprietary and may be subject to frequent changes as the technology evolves. Unfortunately, these bargaining games are not convex and learning is hard in general for non-convex games, but we are still able to show that Follow the Regularized Leader (FTRL) is no regret for our bargaining game.~\footnote{Strategies learned from a no regret algorithm can be considered the ``ground truth'' in online optimization.}  This motivates why agents might use FTRL to learn strategies. We then show the existence of parameter settings of $\ell_1$-regularized FTRL where the algorithm converges to Nash equilibria (NE) with unequal outcomes. Here, the initialization of the algorithm and the choice of how ties are broken in the optimization problem dictate when agents converge to asymmetric NE. Crucially, our agents still have equal merit after learning how to bargain, yet we show it is possible for agents to receive significantly different wages.

In Section~\ref{sec:related_works} we review related work. Sections~\ref{sec:discriminatory_equilibria} and \ref{sec:learning_in_bargaining} give our main results. This paper establishes the importance of understanding non-distributional sources of inequality, particularly from consensus in strategy choice in games. We highlight additional opportunities for future work throughout. All proofs for this paper can be found in the appendix.

\section{Related Work}
\label{sec:related_works}
\subsection{Discrimination Against Agents of Equal Merit}
 In economics, statistical discrimination demonstrates how initially equal levels of merit \textit{become unequal at equilibrium} because one group of workers finds they are less likely to be hired by a firm, so they end up investing less in relevant skill sets~\citep{phelps1972statistical}. Further, taste-based discrimination allows for accurate but discriminatory outcomes among groups of workers with equal merit due to exogenous differences in a firm's utility function for hiring each group due to direct animus~\citep{arrow2015theory}. Instead, we show that discrimination can still arise among agents of equal skills and symmetric utility functions. 
 
Demographic parity is a fairness metric that may find an accurate algorithm to be discriminatory~\citep{kamiran2012data}. However, this metric randomly assigns resources (i.e., positive predictions) so that they are equal among desired groups and \textit{the mechanism that created the inequality is ignored}. 
Other related notions of bias in ML are \textit{historical bias}~\citep{hellstrom2020bias,mehrabi2021survey,rajkomar2018ensuring,roselli2019managing}, \textit{feedback loops}~\citep{adam2020hidden,lum2016predict,malik2020does}, \textit{label bias}~\citep{dai2020label,jiang2020identifying}, and \textit{representational harms}~\citep{abbasi2019fairness,buddemeyer2021words,cheng2023social, curry2020conversational} . Each of these notions is either describing a distributional source or describes a mechanism other than strategy choice to create asymmetric outcomes among agents of equal merit.

\subsection{Discrimination Analysis in Markets}

We build off the work of~\citet{fish2022s} on non-distributional sources of discrimination in a bargaining game, but we provide fully symmetric agents, a stronger notion of equilibrium, and a different non-distributional mechanism: Rather than beliefs about outside options alone, we also look at an agent's ability to make threats during bargaining.

There are several works which highlight undesirable outcomes of bargaining models, including disparities in wages or employment rates~\citep{cavounidis2015discrimination,fang2011theories,fernandez1989striking,hu2017fairness,lax1989commentary,ulph1998labour}. However, these works assume exogenous distributional inequality in the form of disparity in skills, costs across workers, or bargaining power while we show that discrimination arises \textit{even when we control for exogenous differences}.


There are several works that consider bargaining games with multiple equilibria~\citep{dwork2024equilibria,hyde1997multiple,shaked1994opting}, as we do here. Multiple equilibria are useful for demonstrating the existence of discrimination, because it enables the possibility of both an unfair outcome and a better alternative that's still incentive-compatible. However, these works all have some asymmetry between agents including exogenous costs, who gets to propose first, and outside option values~\citep{dwork2024equilibria,hyde1997multiple,shaked1994opting}. Our work directly extends the scenario modeled by ~\citet{ponsati1998rubinstein} by endogenously modeling an agent's outside option, allowing for more than two agents in our market, and most importantly, imposing symmetry between the bargaining agents. Similarly to us,~\citet{agranov2018use} demonstrate the existence of asymmetric outcomes at SPE  with initially symmetric agents. However, their model does not include initially equal bargaining power, including outside options.

Finally,~\citet{kuksov2024endogenous} also considers the question of how an asymmetric equilibrium arises in a symmetric market based on search costs and uncertainty in payoffs that may be extremely high, but we show that inequality still arises when there is a fixed surplus that agents bargain over. 

\subsection{Critiques of Distributional Equality}
There are a few works that consider alternative sources and presentations of discrimination beyond distributive equality. Several works in political philosophy discuss the shortcomings of a purely distributional view of equality and identify relational equality as a way to address the gaps~\citep{anderson1999point,elford2017survey,schemmel2012distributive}. In the literature on machine learning,~\citet{birhane2021algorithmic} draws attention to the complexities of fairness questions and discusses how relational ethics might be a useful framework.~\citet{kasy2021fairness} similarly recognize the failures of current fairness metrics to capture certain forms of discrimination and they model the ways in which power affects the outcome of an algorithm.~\citet{neuhauser2023improving} investigate the effects of homophily preferences during the growth of networks on the visibility of minority populations. However, these works are not focused on modeling particular sources of non-distributional inequality, particularly in markets.

\subsection{Piecewise Linear Optimization}
Our learning setting requires online optimization over piecewise linear utility functions. While there has been lots of work on online optimization for convex functions~\footnote{See \citet{hazan2016introduction} for an introduction to online convex optimization.}, there has been much less work in online piecewise linear function optimization~\citep{balcan2018dispersion,cohen2017online}. The work there have made restrictive assumptions in the adversarial setting. In our work, we focus on bargaining games rather than arbitrary piecewise linear utility functions, but assume an adversary with different assumptions. We show that a simple discretization of our action space leads to a no regret algorithm with respect to the optimal continuous strategy.~\citet{kroer2015discretization} study optimal discretizations of continuous action spaces in extensive form games, but their work does not consider strategies learned under a no-external-regret framework as we do.

\section{Discriminatory Equilibria via Threats}
\label{sec:discriminatory_equilibria}
In this section, we describe a bargaining game where agents are able to use credible threats to set their strategies in a way that produces discriminatory outcomes at equilibrium. This section first sets up our symmetric market, and then characterizes equilibrium outcomes in our first main theorem.

\subsection{The Market Model}

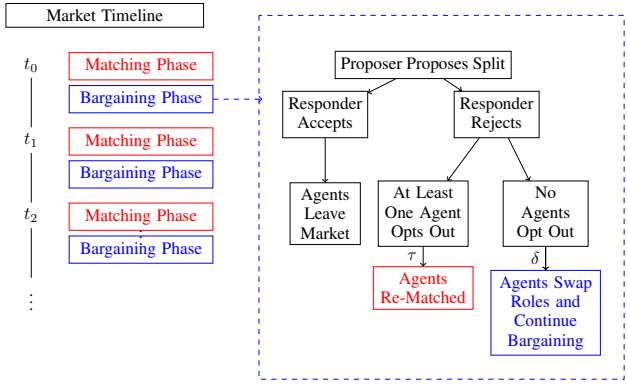
\begin{figure}[t]
    \centering
    \resizebox{\columnwidth}{2in}{
    \begin{tikzpicture}
    \node[rectangle, draw,text width=3.8cm,align=center] (title) at (-6,1) {Market Timeline};
    
    \node (t0)  at (-7.5,0) {$t_0$};
    \node (t1) [below=of t0] {$t_1$} edge (t0);
    \node (t2) [below=of t1] {$t_2$} edge (t1);
    \node (dots)[below=of t2] {$\vdots$} edge (t2);
    
    \node[rectangle, draw,color=blue,text width=2.7cm,align=center] (bargain0) [below right=0.15cm and 0.5cm of t0] {Bargaining Phase};
    \node[rectangle, draw,color=red,text width=2.7cm,align=center] (match0) [above=0.1cm of bargain0] {Matching Phase};
    \node[rectangle, draw,color=blue,text width=2.7cm,align=center] (bargain1) [below right=0.15cm and 0.5cm of t1] {Bargaining Phase};
    \node[rectangle, draw,color=red,text width=2.7cm,align=center] (match1) [above=0.1cm of bargain1] {Matching Phase};
    \node (dots2)[below=0.5cm of bargain1] {$\vdots$};
    
    \node[rectangle, draw,color=blue,text width=2.7cm,align=center] (bargain2) [below right=0.15cm and 0.5cm of t2] {Bargaining Phase};
    \node[rectangle, draw,color=red,text width=2.7cm,align=center] (match2) [above=0.1cm of bargain2] {Matching Phase};

    \node (1b) [right=1cm of bargain0] {};
    \draw[->, dashed, blue] (bargain0) to (1b);

     \node[rectangle, draw] (propose) at (0.5,0) {Proposer Proposes Split};
    \node[rectangle, draw,text width=1.5cm, align=center] (accept) at (-1.5,-1) {Responder Accepts} edge [<-] (propose);
    \node[rectangle, draw,text width=1.5cm, align=center] (reject) at (2,-1)  {Responder Rejects} edge [<-] (propose);
    \node[rectangle, draw,text width=1.2cm, align=center] (leave) at (-1.5, -3) {Agents Leave Market} edge [<-] (accept);
    \node[rectangle, draw,text width=1.6cm, align=center] (optout) at (0.5,-3) {At Least One Agent Opts Out} edge [<-] (reject);
    \node[rectangle, draw,text width=1.5cm, align=center] (nooptout) at (3,-3) {No Agents Opt Out} edge [<-] (reject);
    \node[rectangle, draw, color=red,text width=1.8cm,align=center] (match) at (0.5, -4.5) {Agents \\Re-Matched};
    \node[rectangle, draw, color=blue,text width=2cm, align=center] (bargain2) at (3, -5) {Agents Swap Roles and Continue \\Bargaining} edge [<-] (nooptout);
    \draw[->] (optout) to node[midway,left] {$\tau$}(match);
    \draw[->] (nooptout) to node[midway,left] {$\delta$}(bargain2);

    \node (r1) [right=1cm of title] {};
    \node (r2) [below right=0.5cm of bargain2] {};
    \draw[dashed,blue] (r1) rectangle (r2);

    \end{tikzpicture}
    }
    \caption{The progression of the market over time is shown on the right with one round of the bargaining phase expanded on the left.}
    \label{fig:market_overview}
\end{figure}

Our market is an extensive-form game where candidates and firms are matched and bargain over the split of the surplus generated as a result of the candidate's employment. To control for exogenous differences between agents, we make the normative assumption that all firms and candidates are equally skilled and entitled to the surplus.  The market timeline and the bargaining game can be visualized in Figure~\ref{fig:market_overview} and we go through the details below.

In this market, there are discrete time steps where there is a matching and a bargaining phase. During the matching phase, unmatched firms and candidates are matched with some probability which represents a candidate applying for a job with the firm. We say two agents are the same type if they play the same strategy and assume there is a constant probability of matching with a strategy of each type.

Once two agents are matched, they enter the bargaining phase. Here, firms and candidates participate in an extensive form Rubinstein-style bargaining game~\citep{osborne1990bargaining} to determine the split of the surplus, which we normalize to 1. To ensure agents have a priori equal bargaining power, we insist a newly matched agent gets to propose a split of the surplus first with probability $\frac{1}{2}$ (or the other agent proposes first). 

Given agent $i$ is the proposer and agent $j$ the responder, agent $i$ will propose a split of the surplus $(1-y, y)$ for some $y \in [0,1]$. As the responder, agent $j$ can either reject or accept the proposal. If they accept, they both leave the market. If they reject, Both agents have a chance to opt out of the match if the responder rejects the offer (which without loss of generality can happen simultaneously, as in~\citet{ponsati1998rubinstein}) and otherwise in the next round the proposer becomes the responder and vice versa and a discount factor $\delta$, $0 < \delta < 1$, is incurred. So, if agent $i$'s offer of $y$ is accepted by agent $j$ at round $k \ge 1$, then, agent $i$ gets utility $\delta^k(1-y)$ and agent $j$ gets $\delta^k y$. When either agent chooses to opt out, both agents must pay a cost of $0\le \tau \le 1$ and are sent back into the market pool to be re-matched in some future time. Here, $\tau$ captures the waiting and matching cost incurred when an agent decides to re-enter the market.

The agents' strategies are given as automata with a base state and a threat state. Each state consists of the offer an agent would make in each round as the proposer, the amount they would accept in each round as the responder, and whether they opt out if a proposal is rejected. The agents start using the base state and move to the threat state if their bargaining opponent deviates from the equilibrium path. The strategy table of each agent can be found in the appendix.




For convenience, we make a large market assumption: Our market remains a constant size with the same distribution of firm's and candidate's strategies at all time steps such that the composition of our market remains unchanged over time.  For example, whenever two agents leave the market, we can assume two agents with identical strategies enter in the next time step.

\subsection{Disparity at Equilibrium}
\label{sec:main_theorem}
In this section, we will focus on the case of one type of firm and two types of candidates which we will call $c_1$ and $c_2$ candidates. Since we have only two kinds of bargaining matches, let $p$ be the probability of a firm matching with a $c_1$ candidate and otherwise they match with a $c_2$ candidate. 

We now state our first main theorem which gives the range of payoffs each type of agent can receive at SPE.  Notably, there are SPE where payoffs are asymmetric between firms and candidates and between candidate types.

\begin{customthrm}{1}
    \label{thrm:mult_equilib}

     If $\tau \le \frac{\delta^2}{1+\delta}$, then for any $p\in[0,1]$ and any $w_1,w_2\in[0,1]$ that satisfy
     \begin{align*}
    w_k &\le \frac{1}{2}\left(\frac{1+\delta-2\tau}{1-\tau}\right) \text{ for } k\in\{1,2\}, &(1)\\
    w_1 &\ge \frac{1}{2}\left(\frac{1-\delta+2\tau(1-p)w_2}{1-\tau p}\right), &(2)\\
    \text{and }w_2 &\ge \frac{1}{2}\left(\frac{1-\delta+2\tau p w_1}{1-\tau(1-p)}\right), &(3)
     \end{align*}
     there exists an SPE where the firms obtain an expected payoff of $pw_1+(1-p)w_2$, the $c_1$ candidates get an expected payoff of $1-w_1$ and the $c_2$ candidates get an expected payoff of $1-w_2$ at equilibrium.

\end{customthrm}

 These results hold because there is a range of first round offer values that are immediately accepted as a result of ``threats'' made in the second round where each first round offer in that range corresponds to an SPE, so long as the threats use \textit{the actual outside option of each agent}. So, all that is required is that agents in the market have a consensus in their strategies about each others' outside options to establish the range of strategy profiles in SPE. Such a consensus drives agent acceptance of a discriminatory outcome at equilibrium despite their equal merit.
 
 Though $c_1$ and $c_2$ candidates need not explicitly agree on each other's outside options through the choice of a strategy, there is a dependence between their outside options because the outside option of the firm relative to one group depends on what the other group is willing to accept from or offer to the firm. Hence, Theorem~\ref{thrm:mult_equilib} sets up the bounds of the payoffs to each group of candidates (related to their outside options via $\tau$) in terms of the payoff to the other. Therefore, asymmetric outcomes exist when there are gaps between payoffs that satisfy assumptions (1) - (3) in the theorem statement.
 


\subsection{Discussion}\label{sec:eqlbrm_discussion}

To find gaps in payoffs, first suppose $w_2>w_1$ such that $c_1$ candidates are getting a higher payoff than $c_2$ candidates at equilibrium. Then, we can fix $w_2$ at its upper bound given by assumption (1) and find the difference with the lower bound of $w_1$ given by assumption (2). The largest such gap is given by $\frac{\delta-\tau}{1-\tau p}$. When $\tau = \frac{\delta^2}{1+\delta}$, the gap size grows with $p$ given a fixed $\delta$ and with $\delta$ given a fixed $p$. Notably, a payoff gap persists between $c_1$ and $c_2$ candidates even as $p$ approaches 0. Additionally, even when candidate groups are equally sized (i.e., $p=0.5$), a larger $\delta$ can cause the gap to be larger than $\frac{1}{2}$ such that candidate groups are getting significantly different splits of the surplus.~\footnote{Qualitatively similar results hold for markets with $m$ kinds of firms and $n$ kinds of candidates for any finite $m$ and $n$, and these results can be found in the appendix.}

In this market, no agent has any advantage over the others in terms of information, power, or merit in the bargaining game at the start and yet, as we have shown, it is possible for agents to choose strategies that are in SPE where one type of candidate receives a greater split of the surplus than the other. As such, this market is susceptible to discrimination via credible threats without any initial asymmetric advantage among any of the agents. Although fairness metrics based on distributive equality could detect this kind of discrimination, they would aim to correct the discrimination by equalizing the split of the surplus. However, our model shows that equal entitlement to the surplus did not prevent the discriminatory outcome such that an intervention to equalize resources may not last. 

 
Now consider another case, where $w_1=w_2 = \frac{1}{2}(\frac{1+\delta-2\tau}{1-\tau})$ and this creates a payoff gap of $\frac{\delta-\tau}{1-\tau}$ between the firm and both kinds of candidates. Note that this gap grows with $\delta$ and approaches 1 for a fixed $\tau < 0.5$. Here, the firms' strategy does not depend on the candidate type and the candidates receive the same split of the surplus at the end. However, the firms then get more of the surplus than the candidates even though we assumed that the firms and candidates are equally entitled to the surplus, indicating that the firms were able to acquire more bargaining power than the candidates at equilibrium.


At this point, we have described how our non-distributional source of credible threats in a strategy can produce a number of discriminatory outcomes and power advantages at equilibrium, but we do not say how agents might \textit{learn} to play a strategy in one of these equilibria. So, we turn to learning in bargaining games in Section~\ref{sec:learning_in_bargaining}.  


\section{Discriminatory Equilibria via Learning}
\label{sec:learning_in_bargaining}
In this section, we initiate a study of online learning in bargaining games.  We provide conditions where an online learning algorithm converges to an unequal NE to demonstrate how learning algorithms can produce discriminatory outcomes by learning from the outcome of a repeated game, rather than from distributional inequalities in training data.  Demonstrating this does not require the full model introduced in Section~\ref{sec:discriminatory_equilibria}, so for the sake of simplicity, we use a two-player bargaining game with finitely many rounds of bargaining.  We also motivate the use of this algorithm by showing conditions under which the algorithm is no regret.  This implies agents may want to employ the algorithm regardless of which algorithm the other agent uses. 


\subsection{Simplified Bargaining Game}
In this section, we consider a bargaining game between two agents: $P$ is the agent who proposes first and $R$ is the agent who responds first. Let $\mathcal{G}^{(n)}$ be a bargaining game with $n < \infty$ rounds of bargaining. In each round, a proposer makes an offer in $[0,1]$ to the responder and the responder simultaneously specifies an acceptance threshold in $[0,1]$ indicating the minimum offer they would accept. If the agents don't agree (i.e., the offer is less than the acceptance threshold), the proposer becomes the responder and vice versa in the next round. Agents no longer have an outside option, so if no deal is made within $n$ rounds, both agents get 0. 

We represent the strategy of each agent $i \in\{P,R\}$ as an $n$-dimensional vector $s_i \in [0,1]^n$, representing what they would do at each round of bargaining. Let $s_{i,k}$ be the strategy of agent $i$ at round $k$ of bargaining.  Since two agents alternate being proposer and responder, if $i$ is the proposer at round $k$, $s_{i,k} \in [0,1]$ is an offer to agent $j$ (their opponent) and otherwise $s_{i,k} \in [0,1]$ is an acceptance threshold. Given a strategy profile $(s_P, s_R)$, the utility payoff function to each agent in game $\mathcal{G}^{(n)}$ matches the utility described in Section~\ref{sec:discriminatory_equilibria}, though agents get 0 utility if no offer is accepted at any round.  With this setup, note that $\mathcal{G}^{(1)}$ is the well-studied ultimatum game.~\footnote{Note that for $\mathcal{G}^{(1)}$ any strategy profile in NE is also in SPE. See~\citep{debove2016models,tadelis} for studies of the Ultimatum game.}  

The strategy profiles that are in pure NE are those where $s_{P,k} = s_{R,k}$ for the round $k$ where each agent is getting maximum possible utility, given their opponent's strategy. We will say the \textit{value} of the equilibrium is $s_{P,k}$, i.e., the surplus split that the responder in round $k$ accepts.

\subsection{Learning to Bargain via FTRL}

Learning bargaining strategies in our market is an \textit{online optimization} problem where each agent commits to a strategy at each time step and then learns the utility of their strategy. The standard goal for learning in such settings is to suffer no external regret in the limit, where external regret is the difference in utility between the chosen strategies at each time step and the single best strategy in hindsight given an adversary choosing utility feedback at each time step. An online learning algorithm over a game is \textit{no regret} if its external regret is sublinear in $T$, the number of time steps it runs, for any sequence of adversarial strategies chosen by an opponent~\citep{hazan2016introduction}. This is a very common notion of accuracy in online optimization because in general it's not possible to achieve any stronger notion.

Follow the Regularized Leader (FTRL) is a simple algorithm used in many online \textit{convex} optimization problems that achieves no regret in games when, for example, the utility functions are convex and the strategy space is discrete~\citep{shalev2012online}. We start by motivating the use of FTRL by showing when it is no regret here as well.  However, the utility functions for $\mathcal{G}^{(n)}$ are notably non-convex (or concave), but rather piece-wise linear with jump discontinuities. Moreover, the strategy space is continuous rather than discrete.  So we use Algorithm~\ref{alg:modified_ftrl}, which discretizes the strategy space $[0,1]^n$ before running FTRL.  This ensure that the \textit{expected} utility function is convex with respect to mixed strategies over the discretized space. Then, we can use the previous guarantees of FTRL in online convex optimization to give conditions under which Algorithm~\ref{alg:modified_ftrl} is a no regret learner, even with respect to the optimal strategy in the continuous strategy space.

Algorithm~\ref{alg:modified_ftrl} is parameterized by the number of time steps $T$, a positive integer $D>1$ for discretizing the strategy space, a learning rate $M$, and $p\ge 1$ (since we use the $\ell_p$ norm as the regularizer). At each time step, agent $i$ receives \textit{full feedback} $u_i^{(t)}$, i.e. a vector indexed by $\mathcal{S}^n$ that gives the payoff had agent $i$ played $s\in \mathcal{S}^n$ given their opponent's actual strategy at time $t$. Then, the FTRL update step is used to update agent $i$'s \textit{mixed} strategy based on the expected utility $i$ gets from their strategy at time $t$.

Even though this algorithm always returns a mixed strategy over a finite space with precision $1/D$, we'd like to show that this Algorithm~\ref{alg:modified_ftrl} is a no regret learner with respect to the optimal strategy over the \textit{continuous} space $[0,1]^n$.  However, if the adversary is allowed to specify strategies with arbitrary precision while the learner isn't, the adversary can take advantage of the additional precision to ensure the learning agent gets linear regret. So, we assume the adversary also must discretize $[0,1]$ into bins of size at least $\frac{1}{D}$ before choosing their strategy, though they are allowed to choose a different discretization for each of the $n$ rounds of bargaining. Then, the following proposition establishes sufficient conditions for no regret:

\begin{customprp}{3}
\label{prop:no_regret_result}
In game $\mathcal{G}^{(n)}$, for any agent $i \in \{P,R\}$, when $D = T$, $M= O(1/\sqrt{T})$, and $p=2$, Algorithm~\ref{alg:modified_ftrl} is no regret with respect to the optimal continuous strategy in hindsight for agent $i$.
\end{customprp}

For our NE convergence results, we use $p=1$ for ease of analysis, and we leave for future work extending our convergence results to other choices of regularizer. The use of the $\ell_1$ regularizer implies that each $w_i^{(t)}$ is a pure strategy, which is not the case for other regularizers.
It may also be the case that multiple pure strategies have the same objective value, so in the following analysis, we break ties by choosing the largest tied acceptance threshold for the responder and the largest tied offer for the proposer in each round.
 

\begin{algorithm}
\caption{Discretized FTRL with $\ell_p$ regularizer}\label{alg:modified_ftrl}
\begin{algorithmic}
\STATE \textbf{Input: } $T, M, D, p$, $i \in \{P,R\}$, $\mathcal{G}^{(n)}$
\STATE $\mathcal{S} \gets \{0,\frac{1}{D}, \frac{2}{D}, \ldots, \frac{D-1}{D}, 1\}$
\STATE $w_i^{(1)} \in \Delta(\mathcal{S}^n)$
\COMMENT{\textit{pure} initial strategy}
\STATE $\alpha_i \in \Delta(\mathcal{S}^n)$
\COMMENT{\textit{pure} strategy reference point}
\FOR{$t = 1$ to $T$}
\STATE Play $w_i^{(t)}$ and observe the feedback $u_i^{(t)}$ 
\STATE Update $w_i$: $$w_i^{(t+1)} = \argmax_{w \in \Delta(\mathcal{S}^n)} \langle w, \sum_{\tau=1}^{t}u_i^{(\tau)}\rangle - \frac{\frac{1}{p}\|w -\alpha_i\|_p^p}{M}$$
\ENDFOR
\end{algorithmic}
\end{algorithm}

In the following sections we show (1) convergence to NE occurs when both agents learn to bargain via Algorithm~\ref{alg:modified_ftrl} in $\mathcal{G}^{(1)}$ and at least in some cases in $\mathcal{G}^{(2)}$ and (2) there exist initial conditions such that agents converge to a NE with \textit{asymmetric payoffs}.  Since we have assumed that agents are equally entitled to the surplus here, these results show the importance of understanding how initial conditions and algorithm design choices, like a tie-breaking mechanism, contribute to possibly discriminatory outcomes at NE.

\subsection{Convergence to Nash Equilibrium}
In this subsection, we show a range of conditions under which Algorithm~\ref{alg:modified_ftrl} converges to NE for $n=1$ and $n=2$. We use a strong notion of convergence where there exists a time step $t'$ such that the strategy profiles each agent plays are at NE and that for all $t \ge t'$, they remain at that NE. 

Theorem~\ref{thrm:n=1_convergence} first characterizes convergence in $\mathcal{G}^{(1)}$ which reflects the intuition that a responder should accept any non-zero offer, so the convergence value is the lowest offer the responder sees in the first few time steps. 

\begin{customthrm}{4}
\label{thrm:n=1_convergence}
 For any setting of the initial conditions $w_{r}^{(1)},w_{p}^{(1)},\alpha_r,\alpha_p \in \mathcal{S}\setminus \{0,1\}$, Algorithm~\ref{alg:modified_ftrl}, parameterized by $M > 2D$ and $p=1$, converges to a NE in $\mathcal{G}^{(1)}$ whose value is $\min \{w_{r,1}^{(1)},w_{p,1}^{(1)},\alpha_r\}$.  
\end{customthrm}

However, in games with $n \ge 2$, both agents have a chance to propose and agents can make strategic \textit{threats} about how they will act in the second round of bargaining, which the other agent will observe in this full feedback model. Below we show convergence occurs for at least some initial conditions in the game $\mathcal{G}^{(2)}$. 

\begin{customthrm}{5}
\label{thrm:n=2_convergence}
    Suppose the following initial conditions hold:
    \begin{align*}
        1-w_{r,1}^{(1)} &\ge \delta w_{r,2}^{(1)}, &\\
        w_{p,1}^{(1)} &> \delta(1-w_{p,2}^{(1)}), \text{ and } & \\
        \alpha_{i,1} &> w_{j,1}^{(1)}. &
    \end{align*}
    Then Algorithm~\ref{alg:modified_ftrl}, parameterized by $D>\frac{1}{1-\delta}$, learning rate $M > 2D$, and $p=1$, converges to a NE in $\mathcal{G}^{(2)}$.
\end{customthrm}

The fact that we can characterize convergence in these cases means learning in games $\mathcal{G}^{(n)}$ for $n \ge 2$ may be of independent interest and we leave full analysis for future work. 


\subsection{Disparity at Equilibrium}
\label{sec:4_disparate_outcomes}

We now highlight a range of initial conditions  where agents end up with asymmetric payoffs at equilibrium after learning strategies for $\mathcal{G}^{(1)}$ and $\mathcal{G}^{(2)}$ via Algorithm~\ref{alg:modified_ftrl} with $p=1$. Recall our assumption that all agents are equally entitled to the surplus, so cases with asymmetric payoffs are potentially discriminatory.

First, for $\mathcal{G}^{(1)}$, Theorem~\ref{thrm:n=1_convergence} characterizes a broad range of possible NE convergence values which depend on the initial conditions. Specifically, any case where $\min \{w_{r,1}^{(1)},w_{p,1}^{(1)},\alpha_r\} \neq 0.5$ will converge to a NE with asymmetric payoffs. Next, for $\mathcal{G}^{(2)}$, we start by using Theorem~\ref{thrm:n=2_convergence} to illustrate how it's possible to converge to almost maximally disparate outcomes by only changing the initial conditions: 


\begin{customex}{6}
\label{ex:n=2_disparate_outcomes}
Suppose both a proposer $f$ and responder $c$ use Algorithm 1, parameterized by $D>10, M> 2D, p=1,\delta = 0.9$ to learn in $\mathcal{G}^{(2)}$. If  $w_f^{(1)} = (0.5, w_{f,2}^{(1)} > 1-\frac{0.5}{0.9}), w_{c}^{(1)} = (\frac{1}{D}, 1), $ and $\alpha_{i,1} > w_{j,1}^{(1)} \text{ for } i \in \{f,c\}$, then the algorithm converges to a NE whose value is $\frac{1}{D}$. If, instead, $c$ uses the initial strategy $w_{c'}^{(1)} = (\frac{D-1}{D}, \frac{1}{D\cdot 0.9})$, then the algorithm converges to a NE whose value is $\frac{D-1}{D}$. 
\end{customex}

Suppose there exists one firm $f$ and two candidates, $c_1$ and $c_2$, and each candidate is independently learning to bargain with the firm. Then, Example~\ref{ex:n=2_disparate_outcomes} shows that even if $f$ starts as the proposing agent and offers $0.5$ to both candidates in the first round, then, $c_1$ and $c_2$ can end up in different and asymmetric equilibria with $f$. If $c_1$ uses the strategy $w_c^{(1)}$ they get a payoff of $\frac{1}{D}$ and if $c_2$ uses the strategy $w_{c'}^{(1)}$ they get a payoff of $\frac{D-1}{D}$ for any $D>10$. 

Empirically, we find that $Algorithm~\ref{alg:modified_ftrl}$ converges to NE even outside the conditions of Theorem~\ref{thrm:n=2_convergence}, and are typically asymmetric equilibria.
We implemented Algorithm~\ref{alg:modified_ftrl} using CVXPY~\citep{diamond2016cvxpy} to give empirical convergence results for a range of possible initial conditions. The parameters we use are $T=300, D=16,M=40, p=1$. Since we use $p=1$ for the regularizer, we ensure only pure strategies are chosen at each time step by rounding to 5 decimal places and break ties as described above.  

\begin{figure}
    \centering
    \includegraphics[width=\linewidth]{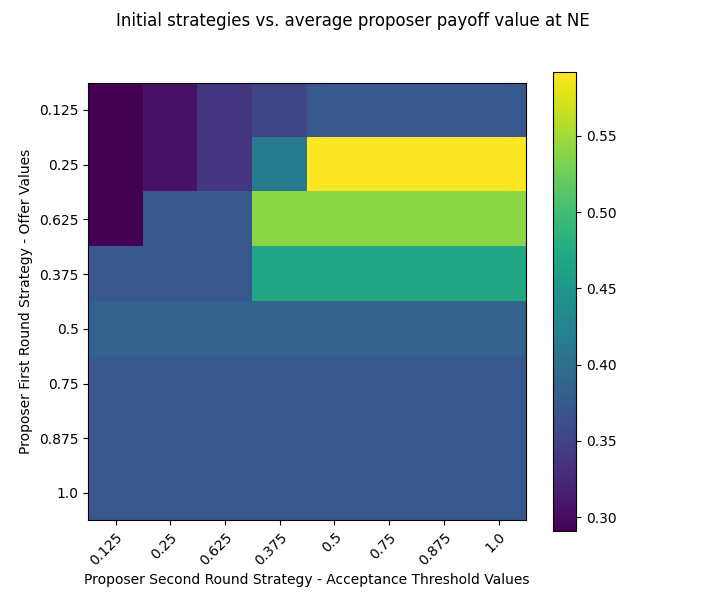}
    \caption{Simulation results for NE outcomes of agents learning strategies in $\mathcal{G}^{(2)}$. The setting is $T=300, M=40, D=16, p=1, \alpha_P = (0.125, 0.375), \alpha_R = (0.375, 0.875)$. The initial strategy of the proposer varies from $\{\frac{1}{D},\frac{3}{D}\ldots, \frac{D-1}{D}\}$ in both strategy dimensions. The color of each cell represents the average payoff to the proposer playing that initial strategy over the initial strategies the responder plays from the same set.}
    \label{fig:G_2_results}
\end{figure}

We first chose a subset of strategy values from $\{0,\frac{1}{D},\ldots,1\}$ and fix each reference point. Then, we ran the algorithm for each possible pairing of initial strategies of the proposer and the responder from this subset. Empirically, we observe that the algorithm converges to NE within $T=300$ time steps for all of these initial conditions, and the color of each cell in Figure~\ref{fig:G_2_results} corresponds to the average payoff to the proposer using a fixed strategy over all possible responder strategies at the final converged NE point. Figure~\ref{fig:G_2_results} shows that high first round initial offers result in a fairly stable payoff for the proposer, low initial first round offers have greater variability. In particular, the proposer gets less than half of the surplus on average if they also have a low initial second round acceptance threshold, but the proposer gets close to or more than half of the surplus if they have a high initial second round acceptance threshold.  Additional simulation results can be found in the appendix.

\subsection{Discussion}
The purpose of this section is to show that online learning algorithms could converge to asymmetric equilibria in bargaining games. First,  Proposition~\ref{prop:no_regret_result} demonstrates that no regret learning is possible for our game which justifies that learning algorithms in this and similar settings are likely to be successful. 

In Theorems~\ref{thrm:n=1_convergence} and~\ref{thrm:n=2_convergence}, we were able to show conditions where Algorithm~\ref{alg:modified_ftrl} converges to a NE and we show \textit{how} the initial conditions and the tie-breaking mechanism determine which equilibrium an algorithm converges to. Then, Example~\ref{ex:n=2_disparate_outcomes} and Figure~\ref{fig:G_2_results} illustrate examples of asymmetric outcomes. It may seem trivial to avoid these cases by setting the relevant parameters carefully. However, our simplified algorithm where we have control over the entire design is not so realistic. Notably, there are possibly analogies to the real-world for why different agents may use different parameters in practice. We can think of $w_f^{(1)}, w_c^{(1)}$ and $
\alpha_f,\alpha_c$ as the ``first offers'' and ``private valuations''of a firm and candidate, respectively. These values possibly depend on things like information about outside options of the bargaining game, aggression of the negotiator, and past bargaining experiences. 

As such, characterizing the convergence of Algorithm~\ref{alg:modified_ftrl} for arbitrary parameter initializations is a start to understanding how to model the complex, possibly non-distributional, considerations of decision makers in bargaining games so that we may understand how to mitigate discriminatory effects.

\bibliography{mybib}

\appendix
\onecolumn

\section{Section 3 Proofs}

We begin by introducing notation and the strategy structure that agents use in bargaining game. During any given bargaining game each agent has an expected payoff of that game which we will call $W_{ij}$ for agent $i$ when playing agent $j$. Note that, $W_{ji} = 1-W_{ij}$ since they are bargaining over a surplus normalized to 1.  We will use $U_i$ as the expected outside option for agent $i$, interpreted as the expected payoff for $i$ at the point in time they opt out of their current negotiation. We only consider strategies where $W_{ij}$, $W_{ji}$, and $U_i$ do not depend on time. 

To compute the outside option of an agent ($U_i$), we multiply the outside option cost $\tau$ with the expected payoff to agent $i$ across all possible agents that $i$ can bargain with which we will notate by $$W_i= \mathbb{E}_{j}[W_{ij}].$$ Thus, $$U_i=\tau W_i.$$

The strategies that we will show are in SPE with these payoffs are given as automata~\citep{osborne1990bargaining} in Table~\ref{table:strats} which are based on strategies defined in \citep{ponsati1998rubinstein} and parameterized by offers $z_{ij}$ that agent $i$ proposes to agent $j$. Fix agent $i$ as an arbitrary agent in the market of some type (a firm or either candidate type). Then, let $\pi_i(Z,U)$ be the strategy of agent $i$ where agent $i$ plays according to Table~\ref{table:strats}, parameterized by $Z=\{z_{ij}\}$ and $U=\{u_{i}\}$.  We will refer to this strategy as $\pi_i$ for brevity in the following Section 3 proofs. 

\begin{table}[h]
\centering
\begin{tabular}{| c| c | c |} 
 \hline
 Agent Actions & Base State & Threat State  \\ [0.5 ex] 
 \hline\hline
 Agent $i$ Propose & $(z_{ij},1-z_{ij})$ & $(1-\frac{u_j}{\delta}, \frac{u_j}{\delta})$ \\  [2.0 ex]
 \hline
 Agent $i$ Accepts & - & Iff $y \geq \frac{u_i}{\delta}$ \\
 \hline
 Agent $j$ Propose & - & $(\frac{u_i}{\delta},1-\frac{u_i}{\delta})$ \\ 
 \hline
 Agent $j$ Accepts & Iff $y \geq 1-z_{ij}$ & Iff $y \geq \frac{u_j}{\delta}$ \\
 \hline
 Agent $i$ Opts Out & Iff $1-y \le z_{ij}$ & \makecell{Iff proposer and\\ $y \ge \frac{u_j}{\delta}$} \\ 
 \hline
 Agent $j$ Opts Out & No & \makecell{Iff proposer and\\ $y \ge \frac{u_i}{\delta}$} \\
 \hline 
     \textit{Transitions} & \makecell{Go to Threat \\if agent $i$ deviates \\in this match} & Absorbing \\
 \hline
\end{tabular}
\caption{Strategies for matched agents $i$ and $j$ when agent $i$ proposes first and agent $j$ responds, parameterized by values $z_{ij}$, $u_i\in[0,1]$, e.g. $(z_{ij},1-z_{ij})$ is the split of the surplus agent $i$ proposes to agent $j$.  Note we use $y$ as a generic variable representing an offer to the responder.}
\label{table:strats}
\end{table}
\begin{customthrm}{1}
    \label{thrm:mult_equilib}

     If $\tau \le \frac{\delta^2}{1+\delta}$, then for any $p\in[0,1]$ and any $w_1,w_2\in[0,1]$ that satisfy
     \begin{align*}
    w_k &\le \frac{1}{2}\left(\frac{1+\delta-2\tau}{1-\tau}\right) \text{ for } k\in\{1,2\}, &(1)\\
    w_1 &\ge \frac{1}{2}\left(\frac{1-\delta+2\tau(1-p)w_2}{1-\tau p}\right), &(2)\\
    \text{and }w_2 &\ge \frac{1}{2}\left(\frac{1-\delta+2\tau p w_1}{1-\tau(1-p)}\right), &(3)
     \end{align*}
     there exists an SPE where the firms obtain an expected payoff of $pw_1+(1-p)w_2$, the $c_1$ candidates get an expected payoff of $1-w_1$ and the $c_2$ candidates get an expected payoff of $1-w_2$ at equilibrium.

\end{customthrm}
The range of payoffs in Theorem \ref{thrm:mult_equilib} is a direct result of the existence of a range of possible proposal values for each $z\in Z$.  This range is a function of $u_i$ and $u_j$, so long as each $u_i=U_i$, and satisfy the conditions given in Proposition~\ref{prop:main} (again adapted from~\citep{ponsati1998rubinstein} for our setting). This range of $z$ values arises from several properties of the bargaining game we imported from~\citep{ponsati1998rubinstein}. The proposer has an advantage and can ask for all of the surplus except for the responder's outside option and this defines the upper bound of their $z$ range. Both agents have the opportunity to opt out and get their outside option at each stage of negotiation, so neither agent can take advantage of the other by forcing them to incur an extra discount factor $\delta$. Thus, the responder can make a credible threat to reject any offer less than their equilibrium share since they can opt out if the proposer does not accept their counteroffer. To keep the proposer from opting out after the first round, the responder must offer them at least what their outside option would be with no time discount in the next time step. Thus, the least the proposer is ever willing to accept is related to the most that the responder could get in the next round and this defines the lower bound of their $z$ range. See \citet{ponsati1998rubinstein} for a more complete discussion of the intuition for how multiple equilibria arise out of these types of strategies. 

The following is a version of the main theorem of~\citet{ponsati1998rubinstein} adapted to our setting.

\begin{customprp}{2}
\label{prop:main}

Given that each agent $i$ plays $\pi_i$ as described above, the strategy profile $\pi = \{\pi_i\}$ is at SPE whenever, for all agents $i$, $u_i = U_i$ and, for all agents $j$ that $i$ can bargain with, $z_{ij} \in [1-\delta(1-\frac{u_i}{\delta}), 1-u_j]$, $u_i \le \delta^2(1-\frac{u_j}{\delta})$, and $u_j \le \delta^2(1-\frac{u_i}{\delta})$. 
 
\end{customprp}

We start with a proof outline that reduces this proposition that the market is at SPE to showing that each possible bargaining match between two agents is at SPE given fixed expected outside options $U_i$ under $\pi$. The proof of SPE for the two agent bargaining game is then given by~\citep{ponsati1998rubinstein}. Finally, we use Proposition~\ref{prop:main} to prove Theorem~\ref{thrm:mult_equilib}.

\begin{proof}[Proof outline]

In the strategy $\pi$, each agent's strategy is described by an automoton, and as such, to show that $\pi$ is an SPE, it suffices to show that are no beneficial ``one-shot'' deviations \cite{osborne1990bargaining}, i.e. no single action an agent may take to improve their expected payoff for every possible state the automota may be in, fixing all other actions.  
Moreover, by the assumption that the market is stationary (any agent leaving the market is replaced with an agent of the same type), the expected outside option for any agent $i$ is stationary as well, implying $U_i$ is well-defined.

In $\pi$, each agent starts in the same state at the beginning of bargaining across all times that they are matched, and actions and payoffs depend only on the other agent they are bargaining with.  Thus the optimal action at any given time, fixing all other actions, depends only on the restriction of the game to a single bargaining match with the same, fixed expected outside options $U_i$.


\end{proof}

We can now return to proving Theorem~\ref{thrm:mult_equilib} by using Proposition~\ref{prop:main}.

\begin{proof}
    Consider an arbitrary $w_1,w_2$ that satisfy conditions (1)-(3) in the theorem statement and suppose $\tau \le \frac{\delta^2}{1+\delta}$. Let $p \in [0,1]$ be the probability of a firm matching with a $c_1$ candidate. Then, we will construct strategies for firms $f$, $c_1$ candidates, and $c_2$ candidates that are in SPE with the desired expected payoffs.

    Each agent will use a strategy given in Table~\ref{table:strats}, parameterized by offers.  It suffices to consider strategies that are only dependent on the three types of agents -- firms, $c_1$ candidates, and $c_2$ candidates -- so that $z_{ij}, u_i, \text{ and } u_j$ only depends on the types of $i$ and $j$. As such, we will refer to the parameters as $z_{fc_k}$, $z_{c_kf}$, $u_f$, and $u_{c_k}$, for $k\in\{1,2\}$.

    First, we set $u_f$ and $u_{c_k}$ so that they will be equal to the expected outside option (discussed in Section 3) of the firms and candidates respectively at equilibrium.  Let $$w_f = p w_1 + (1-p)w_2$$ Note that $w_1, w_2 \in [0,1]$ and so $w_f \in [0,1]$.  Then let
\begin{align*}
    u_f &= \tau w_f,\\
    u_{c_1} &= \tau(1-w_1),\\
    \text { and } u_{c_2} &= \tau(1-w_2).\\
\end{align*}

It follows immediately from these choices, the constraints on $w_1$ and $w_2$, and 
the assumption that $\tau\le \frac{\delta^2}{1+\delta}$ that we have $u_f \le \delta^2(1-\frac{u_{c_1}}{\delta})$, $u_{c_1} \le \delta^2(1-\frac{u_f}{\delta})$, $u_f \le \delta^2(1-\frac{u_{c_2}}{\delta})$, and $u_{c_2} \le \delta^2(1-\frac{u_f}{\delta})$ as needed for Proposition~\ref{prop:main}.

We now need to set $z_{fc_k}$ and $z_{c_kf}$, i.e. $Z$.  In particular, we need $z_{fc_k} \in \left[1-\delta\left(1-\frac{u_f}{\delta}\right), 1-u_{c_k}\right]$ and $z_{c_kf} \in  \left[1-\delta\left(1-\frac{u_{c_k}}{\delta}\right), 1-u_{f}\right]$ to be able to apply Proposition~\ref{prop:main}.  But we also need \[1-w_k = \frac{1}{2}z_{c_kf}+\frac{1}{2}(1-z_{fc_k}),\] because we need $1-w_k$ to be the expected payoff for $c_k$ candidates: From Table~\ref{table:strats}, observe that whatever split ($z_{ij}$, 1-$z_{ij}$) is proposed first, it is always accepted.  And recall that an agent proposes first with probability $\frac{1}{2}$ and responds first with probability $\frac{1}{2}$, so the expected payoff for candidate $c_k$ is exactly $\frac{1}{2}z_{c_kf}+\frac{1}{2}(1-z_{fc_k})$.

Moreover, recall that a firm matches with a $c_1$ candidate with probability $p$ and with a $c_2$ candidate with probability $1-p$, so the expression for the expected payoff to a firm, $w_f$, is 
$$p\left(\frac{1}{2}z_{fc_1}+\frac{1}{2}(1-z_{c_1f})\right) + (1-p)\left(\frac{1}{2}z_{fc_2}+\frac{1}{2}(1-z_{c_2f})\right).$$

To satisfy the expected payoff expressions, it suffices to show that we can choose some $z_{fc_k} \in \left[1-\delta\left(1-\frac{u_f}{\delta}\right), 1-u_{c_k}\right] $ and set $$z_{c_kf} = 1+z_{fc_k}-2w_k$$

Where $z_{c_kf} \in \left[1-\delta\left(1-\frac{u_{c_k}}{\delta}\right), 1-u_f\right]$ as well. To satisfy this range on $z_{c_kf}$, the following bounds on $z_{fc_k}$ results from the setting above
\begin{align*}
   2w_k-\delta\left(1-\frac{u_{c_k}}{\delta}\right)\le z_{fc_k} \le 2w_k-u_f.
\end{align*}

Since we want $z_{fc_k}\in \left[1-\delta\left(1-\frac{u_{f}}{\delta}\right), 1-u_{c_k}\right]$, we require  all of the following constraints to hold
\begin{align*}
     2w_k-\delta\left(1-\frac{u_{c_k}}{\delta}\right) &\le 2w_k-u_f, &(i)\\
     2w_k-\delta\left(1-\frac{u_{c_k}}{\delta}\right) &\le 1-u_{c_k}, &(ii)\\
    1-\delta\left(1-\frac{u_{f}}{\delta}\right) &\le 2w_k-u_f, &(iii)\\
    1-\delta\left(1-\frac{u_{f}}{\delta}\right) &\le  1-u_{c_k}. &(iv)\\
\end{align*}
\noindent Solving ($i$)-($iv$) for bounds on $w_1$ and $w_2$ leads to the following constraints
\begin{align*}
         w_1 &\ge w_2 - \frac{\delta-\tau}{\tau(1-p)}, &(1)\\
    w_2 &\ge w_1 - \frac{\delta-\tau}{\tau p}, &(2)\\
    w_k &\le \frac{1}{2}\left(\frac{1+\delta-2\tau}{1-\tau}\right), &(3)\\
    w_1 &\ge \frac{1}{2}\left(\frac{1-\delta+2\tau(1-p)w_2}{1-\tau p}\right), &(4)\\
    \text{and }w_2 &\ge \frac{1}{2}\left(\frac{1-\delta+2\tau p w_1}{1-\tau(1-p)}\right). &(5)
\end{align*}

Notice that $(1)$ and $(2)$ are always satisfied since $w_k \le 1$ and $\frac{\delta-\tau}{\tau(1-p)} \ge 1$ and $\frac{\delta-\tau}{\tau p}\ge 1$ when $\tau \le \frac{\delta}{2}$ which is true when $\tau \le \frac{\delta^2}{1+\delta}$. So, $(1)$ and $(2)$ are trivially satisfied with $w_k \ge 0$.

Therefore, we need $w_1$ and $w_2$ to satisfy $(3)-(5)$ so that $(i)-(iv)$ are satisfied. Since $w_1$ and $w_2$ are given to satisfy these constraints by the theorem statement, we can conclude there exists  $z_{fc_k} \in \left[1-\delta\left(1-\frac{u_f}{\delta}\right), 1-u_{c_k}\right]$ and $z_{c_kf} \in\left[1-\delta\left(1-\frac{u_{c_k}}{\delta}\right), 1-u_{f}\right]$ with $z_{c_kf} =  1+z_{fc_k}-2w_k$.

As a consequence of the way we set the $Z$ values, the expected payoffs to each agent at equilibrium will be such that all firms get an expected payoff of $W_f = pw_1+(1-p)w_2$ and all $c_k$ candidates get an expected payoff of $W_{c_k} = 1-w_k$. As a result, the expected outside options for each agent will be exactly what we set the parameters as, that is, $u_f = U_f$ and $u_{c_k} = U_{c_k}$.
    
Now we have a set of strategies $\pi$ for every agent with $z_{fc_k} \in [1-\delta\left(1-\frac{u_f}{\delta}\right), 1-u_{c_k}]$, $z_{c_kf} \in [1-\delta\left(1-\frac{u_{c_k}}{\delta}\right), 1-u_f]$, $u_f \le \delta^2(1-\frac{u_{c_k}}{\delta})$ and $u_{c_k} \le \delta^2(1-\frac{u_f}{\delta})$. We also have $u_i = U_i$ for all agents $i$ and, therefore, we can apply Proposition~\ref{prop:main} and say that $\pi$ is in SPE where all firms get $pw_1 + (1-p)w_2$ in expectation and all $c_k$ candidates get $1-w_k$ in expectation and this concludes the proof. 
    
\end{proof}

\subsection{Extending to $m$ Firms and $n$ Candidate Types}
To say that similar qualitative results hold, it suffices to show that the cases we highlighted still happen with multiple firms and multiple candidates. That is, there is a case where at least one candidate is getting less than half of the surplus and the other candidates are getting more than half. Further, there exists a case where the firms are getting more than half and the candidates are all getting less.

First, we state the more general version of Theorem 1 and sketch the proof outline since the proof is essentially the same as the Theorem 1 proof.
\begin{customprp}{1}
    \label{prop:multi_case}
    If $\tau \le \frac{\delta^2}{1+\delta}$, then for any $\textbf{p} = \{p_1,\cdots,p_m\}$ set of probabilities of a candidate matching with an $f_i$ firm for $i \in [m]$ and any $\textbf{q} = \{q_1,\cdots,q_n\}$ set of of probabilities of a firm matching with a $c_j$ candidate for $j \in [n]$ and any $\textbf{w} = \{w_{f_ic_j}\}_{i \in [m], j \in [n]}$ that satisfy,
     \begin{align*}
     w_{f_ic_j} &\le \frac{1}{2(1-\tau p_i)}\left(1+\delta - 2\tau p_i - 2\tau\sum_{i'=1, i'\neq i}^mp_{i'}(1-w_{f_{i'}c_j})\right) & (1)\\
    w_{f_ic_j} &\ge \frac{1}{2(1-\tau q_j)}\left(1-\delta+2\tau\sum_{j'=1, j' \neq j}^nq_{j'}w_{f_ic_{j'}}\right) & (2)\\
     \end{align*}
     there exists an SPE where firm $f_i$ gets an expected payoff of $W_{f_i} = \sum_{j=1}^nq_jw_{f_ic_j}$ and candidate $c_j$ gets an expected payoff of $W_{c_j} = \sum_{i=1}^mp_i(1-w_{f_ic_j})$ at equilibrium.
\end{customprp}
\begin{proof}[sketch]
    The proof is essentially the same as the Theorem 1 proof, except now we use Section~\ref{sec:setting_z} to set $Z$ values that give us the appropriate payoff. Note that we still satisfy the $U$ constraints since $W_{f_i} \le 1$ and $W_{c_j} \le 1$ still.
\end{proof}

Now, we will prove that the constraints in the general case exhibit similar behavior as in the 1-Firm, 2-Candidate case. To do this, we will make use of the following lemmas which make simple observations.
\begin{customlem}{2}
\label{lemma:bounds}
The upper bound constraint $(1)$ in Proposition~\ref{prop:multi_case} is always $> \frac{1}{2}$ and the lower bound constraint $(2)$ in Proposition~\ref{prop:multi_case} is always $< \frac{1}{2}$.  
\end{customlem}
\begin{proof}
    To show that $(1)$ is always greater than $\frac{1}{2}$, it suffices to show $$1+\delta - 2\tau p_i - 2\tau\sum_{i'=1, i'\neq i}^mp_{i'}(1-w_{f_{i'}c_j}) >1-\tau p_i$$
    Since $(1-w_{f_ic_j}) \le 1$ for all $i \in [m],j\in [m]$ and $\sum_{i'=1, i'\neq i}^mp_{i'} = 1-p_i$ by the law of total probability, then we have $$1+\delta - 2\tau p_i - 2\tau\sum_{i'=1, i'\neq i}^mp_{i'}(1-w_{f_{i'}c_j})\ge 1+\delta-2\tau p_i-2\tau(1-p_i)$$ Such that it suffices to show $$1+\delta-2\tau p_i-2\tau(1-p_i) > 1-\tau p_i$$ Which is true when $\tau < \frac{\delta}{2}$ which is satisfied when $\tau \le \frac{\delta^2}{1+\delta}$ and $\delta <1$ (equality when $\delta=1$).\\

    Similarly, to show that $(2)$ is always less than $\frac{1}{2}$, it suffices to show $$1-\delta+2\tau\sum_{j'=1, j' \neq j}^nq_{j'}w_{f_ic_{j'}} < 1-\tau q_j$$ Again, since $w_{f_ic_j} \le 1$ for all $i \in [m],j\in [m]$ and $\sum_{j'=1, j' \neq j}^nq_{j'} = 1-q_j$ by the law of total probability, then we have $$1-\delta+2\tau\sum_{j'=1, j' \neq j}^nq_{j'}w_{f_ic_{j'}} \le 1-\delta+2\tau(1-q_j)$$ Such that it suffices to show $$1-\delta+2\tau(1-q_j) <1-\tau q_j$$ Which is true when $\tau < \frac{\delta}{2}$ which is satisfied when $\tau \le \frac{\delta^2}{1+\delta}$ and $\delta <1$ (equality when $\delta=1$).
\end{proof}
\begin{customlem}{3}
\label{lem:upperbound_1}
    The upper bound constraint $(1)$ in Proposition~\ref{prop:multi_case} is always $\le \frac{1}{2}\left(\frac{1+\delta-2\tau}{1-\tau}\right)$. 
\end{customlem}
\begin{proof}
    Since constraint $(1)$ for an arbitrary $w_{f_ic_j}$ depends on what other firms are getting with candidate $c_j$, we will show a contradiction when one firm tries to get more than $\frac{1}{2}\left(\frac{1+\delta-2\tau}{1-\tau}\right)$ and a contradiction when all firms try to get more than $\frac{1}{2}\left(\frac{1+\delta-2\tau}{1-\tau}\right)$. As a result, we can conclude the upper bound on constraint $(1)$ is $\frac{1}{2}\left(\frac{1+\delta-2\tau}{1-\tau}\right)$.\\
    
    Suppose for some $w_{f_kc_j}$ we have $w_{f_kc_j} = \frac{1}{2}\left(\frac{1+\delta-2\tau}{1-\tau}\right)+c$ for some $c >0$. Then, by constraint $(1)$, for all $i \in [m], i\neq k$ we have $$w_{f_ic_j} \le \frac{1}{2(1-\tau p_i)}\left(1+\delta - 2\tau p_i - 2\tau\sum_{i'=1, i'\neq i, i'\neq k}^mp_{i'}(1-w_{f_{i'}c_j}) - 2\tau p_k\left(\frac{1}{2}\left(\frac{1-\delta}{1-\tau}\right)-c\right)\right)$$ 
    Notice that the upper bound is maximized when each $w_{f_{i'}c_j}$ is as large as possible. So, setting $w_{f_{i'}c_j} = \frac{1}{2}\left(\frac{1+\delta-2\tau}{1-\tau}\right)$ for all $i' \neq i, k$ gives an upper bound of $$w_{f_ic_j} \le \frac{1}{2}\left(\frac{1+\delta-2\tau}{1-\tau}\right) + \frac{\tau p_{k}}{1-\tau p_i}c$$ Letting $w_{f_ic_j}$ equal its upper bound gives an upper bound on $w_{f_kc_j}$ of:
    $$w_{f_kc_j} \le \frac{1}{2}\left(\frac{1+\delta-2\tau}{1-\tau}\right) + \frac{\tau p_i \tau p_{k}}{(1-\tau p_i)(1-\tau p_{k})}c$$
    Notice that $\frac{\tau p_i \tau p_{k}}{(1-\tau p_i)(1-\tau p_{k})} < 1$ when $\tau \le \frac{\delta^2}{1+\delta}$ and $p_i <1$ and $p_{k} <1$ which we can assume when there are at least two firms. Therefore, this upper bound contradicts our original setting of $w_{f_kc_j} = \frac{1}{2}\left(\frac{1+\delta-2\tau}{1-\tau}\right)+c$ for some $c >0$. \\

    Suppose now that all firms have $w_{f_ic_j} = \frac{1}{2}\left(\frac{1+\delta-2\tau}{1-\tau}\right)+c$ for some $c >0$. Then, by constraint $(1)$ for all $i \in [m]$, we have $$w_{f_ic_j} \le \frac{1}{2(1-\tau p_i)}\left(1+\delta - 2\tau p_i +2\tau\sum_{i'=1, i'\neq i,}^mp_{i'}\frac{1}{2}\left(\frac{1-\delta}{1-\tau}\right)-2\tau(1-p_i)c\right)$$ Which reduces to $$w_{f_ic_j} \le \frac{1}{2}\left(\frac{1+\delta-2\tau}{1-\tau}\right)+\frac{\tau(1-p_i)}{1-\tau p_i}c$$ Where $\frac{\tau(1-p_i)}{1-\tau p_i} <1$ whenever $\tau <1$. Thus, this upper bound contradicts our original setting of $w_{f_ic_j} = \frac{1}{2}\left(\frac{1+\delta-2\tau}{1-\tau}\right)+c$.
\end{proof}

\begin{customlem}{4}
    \label{lem:lowerbound_2}
    The lower bound constraint $(2)$ in Proposition~\ref{prop:multi_case} is always $\ge \frac{1}{2}\left(\frac{1-\delta}{1-\tau}\right)$.
\end{customlem}
\begin{proof}
    In this case, the lower bound $(2)$ for an arbitrary $w_{f_ic_j}$ depends on what the firm $f_i$ gets with other candidates. Similar to the proof of Lemma~\ref{lem:upperbound_1}, we can show contradictions when firm $f_i$ gets less than $\frac{1}{2}\left(\frac{1-\delta}{1-\tau}\right)$ with one candidate $c_k$ as well as when $f_i$ gets less than $\frac{1}{2}\left(\frac{1-\delta}{1-\tau}\right)$ with all candidates.\\

    Suppose for some $w_{f_ic_k}$ we have $w_{f_ic_k} = \frac{1}{2}\left(\frac{1-\delta}{1-\tau}\right)-c$ for some $c >0$. Then, by constraint $(2)$, we have $$w_{f_ic_j} \ge \frac{1}{2(1-\tau q_j)}\left(1-\delta+2\tau\sum_{j'=1, j' \neq j,j'\neq k}^nq_{j'}w_{f_ic_{j'}} + 2\tau q_k\left(\frac{1}{2}\left(\frac{1-\delta}{1-\tau}\right)-c\right)\right)$$ For $j' \neq j,k$, set $w_{f_ic_{j'}} = \frac{1}{2}\left(\frac{1-\delta}{1-\tau}\right)$ to minimize the lower bound. Then, we have $$w_{f_ic_j} \ge \frac{1}{2}\left(\frac{1-\delta}{1-\tau}\right) -\frac{\tau q_k}{1-\tau q_j}c$$ Letting $w_{f_ic_j}$ equal its lower bound gives a lower bound on $w_{f_ic_k}$ of: $$w_{f_ic_k} \ge \frac{1}{2}\left(\frac{1-\delta}{1-\tau}\right)-\frac{\tau q_j \tau q_k}{(1-\tau q_j)(1-\tau q_k)}c$$ As before, notice that $\frac{\tau q_j \tau q_{k}}{(1-\tau q_j)(1-\tau q_{k})} < 1$ when $\tau \le \frac{\delta^2}{1+\delta}$ and $q_j <1$ and $q_{k} <1$ which we can assume when there are at least two candidates. Therefore, this upper bound contradicts our original setting of $w_{f_ic_k} = \frac{1}{2}\left(\frac{1-\delta}{1-\tau}\right)-c$ for some $c >0$.\\

    Suppose now that firm $i$ has $w_{f_ic_j} = \frac{1}{2}\left(\frac{1-\delta}{1-\tau}\right)-c$ with all candidates $c_j$. Then, by constraint $(2)$, we have $$w_{f_ic_j} \ge \frac{1}{2(1-\tau q_j)}\left(1-\delta+2\tau\sum_{j'=1, j' \neq j}^nq_{j'}\left(\frac{1}{2}\left(\frac{1-\delta}{1-\tau}\right)-c\right)\right)$$ Which reduces to $$w_{f_ic_j} \ge \frac{1}{2}\left(\frac{1-\delta}{1-\tau}\right) - \frac{\tau (1-q_j)}{1-\tau q_j} c$$ Where, again, $\frac{\tau (1-q_j)}{1-\tau q_j} <1$ whenever $\tau <1$. Thus, this lower bound contradicts our original setting of $w_{f_ic_j} = \frac{1}{2}\left(\frac{1-\delta}{1-\tau}\right)-c$. 
\end{proof}
\begin{customprp}{5}
    Constraints $(1)$ and $(2)$ in Proposition~\ref{prop:multi_case} allow for the case where one type of candidate gets significantly less than the other candidates at equilibrium.
\end{customprp}
\begin{proof}
    Because Lemma~\ref{lemma:bounds} does not depend on any $w_{f_ic_j}$ values, we can say there exist $L <\frac{1}{2}$ and $G>\frac{1}{2}$ (which will depend on $\textbf{p}$ and $\textbf{q}$) such that we can pick $\textbf{w}$ values from the range $[L,G]$ and constraints $(1)$ and $(2)$ will always be satisfied.\\

    So, we can pick some candidate index $d \in [n]$ that will be discriminated against. Then, we can choose any $w_{f_ic_d} \in (\frac{1}{2},G]$ for all $i \in[m]$ and any $w_{f_ic_j} \in [L,\frac{1}{2})$ for all $i \in [m], j \in [n], j\neq d$ to ensure that candidate $d$ gets strictly less than $\frac{1}{2}$ while all the other candidates get strictly more $\frac{1}{2}$. Therefore, the constraints allow for a case where one type of candidate gets significantly less than the other candidates at equilibrium. 
\end{proof}
\begin{customprp}{6}
    Constraints $(1)$ and $(2)$ in Proposition~\ref{prop:multi_case} allow for the case where all the firms get more than $\frac{1}{2}$ and all the candidates get less than $\frac{1}{2}$ at equilibrium.
\end{customprp}
\begin{proof}
    Similar to above, we can choose any $w_{f_ic_j} \in (\frac{1}{2},G]$ for all $i \in [m], j\in [n]$ and it will be the case that all firms get more than $\frac{1}{2}$ in expectation and all the candidates get less than $\frac{1}{2}$ in expectation at equilibrium. 
\end{proof}
\newpage

\section{Section 4 Proofs}

\subsection{No Regret Proof}
\begin{customprp}{3}
\label{prop:no_regret_result}
In game $\mathcal{G}^{(n)}$, for any agent $i \in \{P,R\}$, when $D = T$, $M= O(1/\sqrt{T})$, and $p=2$, Algorithm~\ref{alg:modified_ftrl} is no regret with respect to the optimal continuous strategy in hindsight for agent $i$.
\end{customprp}

\begin{proof}
    To begin, let $y^{(t)} = (y_{-i,1}^{(t)}, \ldots, y_{-i,n}^{(t)})$ be the adversarial chosen strategy at time $t$ where the adversary is acting like agent $-i$. Since we are establishing no regret with respect to the optimal strategy in the \textit{continuous} space in hindsight, it suffices to consider an adversary that chooses pure strategies from the space $[0,1]^n$. The only assumption we make is that the adversary also must discretize $[0,1]$ into bins of size at least $\frac{1}{D}$ before choosing their strategy, though they are allowed to choose a different discretization for each of the $n$ rounds of bargaining.~\footnote{It is trivial for the adversary to ensure the learning agent gets linear regret if they are allowed to specify strategy values with arbitrary precision.} Next, let $s_i^* = (s_{i,1}^*, s_{i,2}^*, \ldots, s_{i,n}^*)$ be the optimal continuous strategy after time $T$ for agent $i$ such that  
    $$s_i^* = \argmax_{s \in [0,1]^n} \sum_{t=1}^T u_i(s, y^{(t)}).$$
    Next, let $w_i^* = (w_{i,1}^*, w_{i,2}^*, \ldots, w_{i,n}^*)$ be the optimal discrete strategy after time $T$ for agent $i$ such that  
    $$w_i^* = \argmax_{s \in \Delta(\mathcal{S}_i^n)} \sum_{t=1}^T u_i(s, y^{(t)}).$$
    
    Let $\mathcal{K}_p \subseteq [n]$ be the set of rounds where agent $i$ proposes an offer and let $\mathcal{K}_r\subseteq [n]$ be the set of rounds where agent $i$ responds to an offer. Then, Let $j_k\in [1,D]$ be such that $s_{i,k} \in [\frac{j_k}{D},\frac{j_k+1}{D}]$, and let $a_{i}$ be a pure strategy where for each round $k \in [n]$, $$a_{i,k} = \begin{cases}
        \frac{j_k+1}{D} & k \in \mathcal{K}_p \\
        \frac{j_k}{D} & k \in \mathcal{K}_r\\
    \end{cases}$$

    Then, by the optimality of $w_i^*$ and the fact that $a_{i} \in \Delta(\mathcal{S}_i^n)$, it is the case that
    $$\sum_{t=1}^T  u_i(w_i^*, y^{(t)}) \ge \sum_{t=1}^T u_i(a_{i}, y^{(t)}).$$

    Next, let $N_{s_k}^{(t)}$ be the number of times an offer of $s_k$ gets accepted in round $k$ through time $t$, given the adversarial sequence of feedback functions. That is,
    $$N_{s_k}^{(t)} = \sum_{\tau=1}^t \mathbf{1}\{s_k \ge y_{r,k}^{(\tau)}\}.$$
    Further, let $U_{s_k}^{(t)}$ be the utility that an acceptance threshold of $s_k$ gets in round $k$ through time $t$, given the adversarial sequence of feedback functions. That is, 
    $$U_{s_k}^{(t)} = \sum_{\tau=1}^t \begin{cases}
        \delta^k \cdot y_{r,k}^{(\tau)} & s_k \le y_{r,k}^{(\tau)}\\
        0 & \text{otherwise}
    \end{cases} $$

     By our assumption on the adversary, each unique value in the adversarial strategy $y_{-i,k}^{(t)}$ is at least $\frac{1}{D}$ distance away from any other unique strategy values at round $k$, it will be the case that the learning agent can slightly over- and under-estimate $s_i^*$  to get no regret.

     Specifically, for all $k \in \mathcal{K}^p$, it is true that $$N_{s_{i,k}^*}^{(T)} = N_{a_{i,k}}^{(T)}$$
     because whenever $s_{i,k}^* \ge y_{-i,k}^{(t)}$, then $a_{i,k} \ge y_{-i,k}^{(t)}$ since $a_{i,k} \ge s_{i,k}^*$ and whenever $s_{i,k}^* < y_{-i,k}^{(t)}$, then $a_{i,k} < y_{-i,k}^{(t)}$ by the $\frac{1}{D}$ spacing of each $y_{i,k}^{(t)}$. 

     Further, for all $k \in \mathcal{K}_r$, it is true that $$U_{s_{i,k}^*}^{(T)} = U_{a_{i,k}}^{(T)}$$ because $a_{i,k} \le s_{i,k}^*$, so the strategy  $a_i$ gets utility at round $k$ from each adversarial offer $y_{i,k}^{(t)}$ that $s_i^*$ does. Again, by the $\frac{1}{D}$ spacing of each $y_{i,k}^{(t)}$, the strategy $a_i$ does not get utility in round $k$ from an adversarial offer $y_{i,k}^{(t)}$ that $s_{i,k}^*$ does not get utility from.
     
     Therefore, we can show that the optimal discrete strategy $w_i^*$ gets utility that is within a constant of the optimal continuous strategy $s_i^*$.
    \begin{align*}
        \sum_{t=1}^T\left( u_i(s_i^*, y^{(t)}) -  u_i(w_i^*,y^{(t)})\right) 
        &\le  \sum_{t=1}^T( u_i(s_i^*, y^{(t)}) - u_i(a_{i}, y^{(t)}))\\
    &= \sum_{k \in \mathcal{K}_p}(1-s_{i,k}^*)N_{s_{i,k}^*}^{(T)} - \sum_{k \in \mathcal{K}_p}(1-a_{i,k})N_{a_{i,k}}^{(T)}\\
        \intertext{Substituting in the value of $a_{i,k}$ for $k \in \mathcal{K}_p$ and since $N_{s_{i,k}^*}^{(T)} = N_{a_{i,k}}^{(T)}:$}
        \sum_{t=1}^T\left( u_i(s_i^*, y^{(t)}) -  u_i(w_i^*,y^{(t)})\right) 
        &\le \sum_{k \in \mathcal{K}_p}N_{s_{i,k}^*}^{(T)}\left(\frac{j+1}{D}-s_{i,k}^*\right)\\
        \intertext{Since $D = T$ and $\frac{j+1}{D} -s_{i,k}^* \le \frac{1}{D}$, then this sum is bounded by:}
        \sum_{t=1}^T\left( u_i(s_i^*, y^{(t)}) -  u_i(w_i^*,y^{(t)})\right) 
        &\le \sum_{k \in \mathcal{K}_p}N_{s_{i,k}^*}^{(T)} \cdot \frac{1}{T}\\
        \intertext{Since the number of offers accepted at round $k$ is at most $T$, then we can conclude:}
        \sum_{t=1}^T \left(u_i(s_i^*, y^{(t)}) - u_i(w_i^*,y^{(t)})\right) &\le O(1)
    \end{align*}

    Since Algorithm~\ref{alg:modified_ftrl} with $p=2$ for the reguularizer is no regret with respect to the discretized strategy space when $M = O(\frac{1}{\sqrt{T}})$ (See~\citet{hazan2016introduction} for one such proof), then, for online strategies $w_i^{(t)} $, we will have $$\sum_{t=1}^T u_i(w_i^*,y^{(t)}) -  \sum_{t=1}^T u_i(w_i^{(t)},y^{(t)}) \le O(\sqrt{T}).$$ 
    Thus, agent $i$ learning via Algorithm~\ref{alg:modified_ftrl} gets no regret with respect to $s_i^*$:
    \begin{align*}
        \sum_{t=1}^T\left( u_i(s_i^*, y^{(t)}) - u_i(w_i^{(t)},y^{(t)})\right) &=\sum_{t=1}^T \left(u_i(w_i^*, y^{(t)})-  u_i(w_i^{(t)}, y^{(t)})\right) + \sum_{t=1}^T \left(u_i(s_i^*, y^{(t)}) -  u_i(w_i^*,y^{(t)})\right)\\
        &\le \sum_{t=1}^T \left(u_i(w_i^*,y^{(t)}) -  u_i(w_i^{(t)},y^t)\right)+ O(1)\\
        &\le O(\sqrt{T}) + O(1)\\
        &= O(\sqrt{T})
    \end{align*}
\end{proof}

\subsection{Convergence Proofs}
We use the $\ell_1$ regularizer in Algorithm~\ref{alg:modified_ftrl} throughout this section, which allows us to make a simplification. Specifically, since the initial conditions $w_i^{(1)}$ and $\alpha_i$ are pure strategies, it will never be the case that a mixed strategy gets a strictly higher objective value than a pure strategy since it turns out a mixed strategy's objective value is a convex combination over the objective values of the pure strategies in its support. Therefore, all $w_i^{(t)}$ will be pure. As such, we will refer to $w_i^{(t)}$ as a pure strategy in the subsequent analysis such that $w_i^{(t)}\in \mathcal{S}^n$.
\begin{customthrm}{4}
\label{thrm:n=1_convergence}
 For any setting of the initial conditions $w_{r}^{(1)},w_{p}^{(1)},\alpha_r,\alpha_p \in \mathcal{S}\setminus \{0,1\}$, Algorithm~\ref{alg:modified_ftrl}, parameterized by $M > 2D$ and $p=1$, converges to a NE in $\mathcal{G}^{(1)}$ whose value is $\min \{w_{r,1}^{(1)},w_{p,1}^{(1)},\alpha_r\}$.  
\end{customthrm}

To begin, we make use of the following lemmas.

\begin{customlem}{1}    \label{lem:maximizers}
    For agent $i$, $i\in \{P,R\}$,  updating their strategy via Algorithm~\ref{alg:modified_ftrl} parameterized by $M > 1$, their updated strategy $w_i^{(t)}$ is one of $\{\alpha_i, w_{j}^{(1:t-1)}\}$.
\end{customlem}
\begin{proof}

To begin, note that agents learning according to Algorithm~\ref{alg:modified_ftrl} start with an initial strategy that is pure and they use a pure strategy as a reference point in the regularizer. So, we start by arguing why all maximizers will be pure strategies as well. 

For either agent, given that the other has played only pure strategies at each round, a non-pure mixed strategy will only be a maximzier if there are multiple pure strategies that have equal objective values. This is because  a convex combination of multiple strategies cannot attain a higher objective value than the maximum objective value of any of the individual strategies. So, any mixture of strategies that attain equal objective values will tie as being a possible maximizer. But, in Algorithm~\ref{alg:modified_ftrl}, we break ties by choosing the pure strategy that has the largest support value among all the tied mixed strategy maximizers. Thus, any maximizer will necessarily be a pure strategy. For this reason, we treat $w_i$ and $w_{j}$ as pure strategies for the remainder of the proof. 

We continue by analyzing the structure of our utility functions to find the possible maximizers. The sum of utilities up to time $t$ is a piecewise linear function: Each $u_i(\cdot,w_{j}^{(\tau)})$ given agent $i$ and $\tau\in[t]$ is a piecewise linear function with exactly one discontinuity at the point $w_{j}^{(\tau)}$, and so their sum must be piecewise linear as well with at most $t-1$ jump discontinuities (one for each $w_{j}^{(\tau)}$, though they need not all be unique). The regularizer is also piecewise linear with one jump discontinuity at $\alpha_i$ since any pure strategy not equal to $\alpha_i$ incurs a regularization penalty of $\frac{2}{M}$ (i.e., this value is deducted from their utility sum) while $\alpha_i$ incurs a regularization penalty of 0. Therefore, the entire objective function will be a piecewise linear function with possible discontinuities at $\alpha_i$ and $w_{j}^{(\tau)}$ for all $\tau\in[t]$. 
Moreover, $w_i^{(t)}$ must be the maximum of the maximum of each piece, and so it suffices to show that the slope of each piece for agent $P$ is non-zero and zero for agent $R$, but we break ties as described above. 

First consider agent $P$. Let $w \in \mathcal{S}$ be a pure straetgy, then each piece with $w \neq \alpha_p$ is the sum of utility functions, all with slope of $-1$, and the regularizer deducts a constant of $\frac{-2}{M}$, and thus the objective function must have negative slope for these points. Otherwise, for $w=\alpha_p$, the regularizer is now 0, so $w=\alpha_p$ is a singleton piece. Thus, the maximum of each piece will occur at the lower bound of each piece which are exactly the acceptance thresholds of agent $R$ and $\alpha_p$, so we can conclude $w_p^{(t)} \in \{\alpha_p, w_r^{(1:t-1)}\}$.

Next, consider agent $R$. Here, the only possible maximizers are strategies $w \le \alpha_r$. This is because each strategy is an acceptance threshold and, thus, gets utility for all $P$ offers that are at least equal to this threshold. So, smaller acceptance thresholds will necessarily get more utility than higher ones while both being equally penalized by the regularizer. Thus, we can ignore pieces with $w > \alpha_r$ to find possible values of $w_r^{(t)}$. Next, each piece with $w < \alpha_r$ is the sum of utility functions with slope 0 and the regularizer which deducts a constant value of $\frac{2}{M}$. So, the sum of slopes will still be 0 and, thus, we break ties in objective values according to Algorithm~\ref{alg:modified_ftrl} by choosing the maximum of each piece. Otherwise if $w = \alpha_r$, the objective value is only the sum of utilities for an acceptance threshold of $\alpha_r$ with no regularization penalty, so $w=\alpha_r$ is a singleton piece. Thus, the maximum of each piece will then be exactly the offers of agent $P$ or $\alpha_r$, so we can conclude $w_r^{(t)} \in \{\alpha_r, w_p^{(1:t-1)}\}$.
\end{proof}
\begin{customlem}{2}
    \label{lem:ref_point_influence}
    If Algorithm~\ref{alg:modified_ftrl} is parameterized by a discretization constant $D >1$ and a learning rate constant $M > 2D$, then each agent will always choose the strategy that maximizes utility in hindsight at each time step and will either break ties by the tie-breaking mechanism or the reference point in their regularization function.
\end{customlem}
\begin{proof}
     To begin, consider the update step for agent $i$ for $i \in \{P, R\}$. At time $t$, agent $i$ chooses their next strategy $w_i^{(t)}$ according to 
    $$w_{i}^{(t)} = \argmax_{w \in \Delta(\mathcal{S}_i^n)} \sum_{\tau=1}^{t-1} u_i(w, w_{j}^{(\tau)}) - \frac{\|w-\alpha_i\|_1}{M}.$$
    Since the agents play a simultaneous bargaining game, we consider mixed strategies over the Cartesian product of the strategy space of $n$ rounds of bargaining for any $n\ge 1$. So, we let $\alpha_i$ represent a vector with a 1 in the dimension corresponding to the strategy $\alpha_i$ and 0 everywhere else. Since agents start with pure strategies initially (and the reference points are pure strategies), then they will continue to choose pure strategies since a convex combination of strategies cannot get more utility than the strategy that gets the maximum utility, given your opponent plays only pure stratgies. Then, for any pure strategy $w \in \Delta(\mathcal{S}_i^n)$, it is the case that $$\|w-\alpha_i\|_1 =\begin{cases}
         2 & w \neq \alpha_i\\
        0 & w = \alpha_i
    \end{cases}$$
    Further, note that the discretization in Algorithm~\ref{alg:modified_ftrl} sets points a distance of $\frac{1}{D}$ apart. By the structure of the utility functions, for the proposing agent in a deal, an offer that is $\frac{1}{D}$ larger (or smaller) than another offer results in that agent getting $\frac{1}{D}$ less (or more) utility and, for the responding agent in a deal, an acceptance threshold that is $\frac{1}{D}$ smaller (or larger) than another acceptance threshold gets the same amount of (or possibly 0) utility. Therefore, the utility-maximizing strategy gets at least $\frac{1}{D}$ more utility than any other non-utility maximizing strategy. 

    Let $\mathcal{U}^{(t)}(w)$ be the utility sum for a strategy $w \in \Delta(\mathcal{S}_i^n)$ at time $t$:
    $$\mathcal{U}^{(t)}(w) = \sum_{\tau=1}^t u_i(w, w_{j}^{(\tau)}).$$

    Let $w_i^* \in \Delta(\mathcal{S}_i^n)$ be the utility-maximizing strategy at $t$ for agent $i$ that would be chosen by the tie-breaking mechanism. Then $w_i^*$ is trivially objective-maximizing compared to any other $w_i \neq \alpha_i$. Further, if $\mathcal{U}^{(t)}(w_i^*) > \mathcal{U}^{(t)}(\alpha_i)$ (which implies $w_i^* \neq \alpha_i$), then $w_i^*$ is objective-maximizing compared to $\alpha_i$ when
    \begin{align*}
        \mathcal{U}^{(t)}(w_i^*) - \frac{2}{M} &>  \mathcal{U}^{(t)}(w_i)\\
        M &> \left(\frac{2}{\mathcal{U}^{(t)}(w_i^*) - \mathcal{U}^{(t)}(w_i)}\right)^2\\
        \intertext{Since it is the case that $\mathcal{U}^{(t)}(w_i^*) - \mathcal{U}^{(t)}(w_i) \ge \frac{1}{D}$, then it suffices to set}
        M &> \frac{2}{\frac{1}{D}}  = 2D. \\
    \end{align*}
    Otherwise, if $\mathcal{U}^{(t)}(w_i^*) \le \mathcal{U}^{(t)}(\alpha_i)$ and $w_i^* \neq \alpha_i$, then $\alpha_i$ is trivially the objective-maximizing strategy since $\frac{2}{M} > 0$ for all $M>0$. So, by setting $M > 2D$, we ensure that the reference point $\alpha_i$ always has a smaller objective value than the utility-maximizing strategy and is not chosen over such a strategy. 
    
    With this setting, the reference point $\alpha_i$ has \textit{weak} influence over the agent's choice of strategy where an agent will only choose $\alpha_i$ when it is utility-maximizing and breaks ties when there are possibly many utility-maximizing strategies.
\end{proof}
We now return to the Theorem 4 proof.
\begin{proof}
To begin, we will refer to the values of $\{w_{p,1}^{(1)}, w_{r,1}^{(1)},\alpha_p,\alpha_r\}$ as the \textit{initial conditions} of Algorithm~\ref{alg:modified_ftrl}.

Next, this proof involves the evaluation and comparison of the objective function of Algorithm~\ref{alg:modified_ftrl} for different pure strategies and by lemma~\ref{lem:ref_point_influence}, the strategy that is utility-maximizing in hindsight will be chosen each round when $M>2D$. So, define $U_i^{(t)}(s_i, s_{j}^{(1:t-1)})$ as the sum of utility agent $i$ gets up until time $t$ given they play a pure strategy $s_i \in \mathcal{S}$ and $ s_{j}^{(1:t-1)} = \{s_{j}^{(1)}, \ldots,s_{j}^{(t-1)}\}$ is the sequence of previous pure strategies agent $j$ has played from time $1$ to $t-1$:
$$U_i^{(t)}(s_i,s_{j}^{(1:t-1)}) = \sum_{\tau =1}^{t-1} u_i(s_i, s_{j}^{(\tau)}).$$

In this proof, we will be able to track the \textit{trajectory} of Algorithm~\ref{alg:modified_ftrl}. Here, the trajectory is the sequence of strategies each agent chooses at each time step given the initial conditions of Algorithm~\ref{alg:modified_ftrl} and the choice of learning rate $M$. 

Using the trajectory of Algorithm~\ref{alg:modified_ftrl}, we will be able to show that two agents learning with this algorithm will \textit{converge} to a Nash Equilibrium (NE) in $\mathcal{G}^1$ whose value is a function of the initial conditions. By converge, we mean there exists a time $t'$ such that for all time steps $t \ge t'$, we have $$w_p^{(t)} = w_r^{(t)}.$$

Specifically, there are two possible initial conditions that lead to different trajectories and convergence times $t'$, but each leads to the same convergence NE value. First, let $$p_{\max} = \max\{\min\{\alpha_r,w_{p,1}^{(1)}\}, w_{r,1}^{(1)}\}$$ and $$p_{\min}=  \min\{\min\{\alpha_r,w_{p,1}^{(1)}\},w_{r,1}^{(1)}\}.$$ 


We will say $\mathcal{C}_1$ and $\mathcal{C}_2$ are each possible initial condition and we define them in disjunctive normal form for convenience. Let $A,B,C, D$ be literals defined as 
\begin{align*}
    A &:= p_{\min} = p_{\max},\\
    B &:= 1-p_{\min} > s(1-p_{\max}),\\
    C &:= \alpha_p  = p_{\min},\\
    D &:= 1-p_{\min} = 2(1-p_{\max}).\\
\end{align*}
Then, $\mathcal{C}_1$ is defined as
$$\mathcal{C}_1 = A \vee (\neg A \wedge B) \vee (\neg A \wedge C\wedge D),$$

and the definition of $\mathcal{C}_2$ follows as $$\mathcal{C}_2 = \neg \mathcal{C}_1.$$




We will now proceed to prove the theorem statement by distinguishing between the two trajectories:
\begin{itemize}
    \item If $\mathcal{C}_1$ is true, then running Algorithm~\ref{alg:modified_ftrl} will result in the trajectory $\mathcal{T}_1$ given by Table~\ref{tab:algorithm_behavior_t_3} where convergence occurs at $t_1' = 3$.
    \item  If instead $\mathcal{C}_2$ is true, then running Algorithm~\ref{alg:modified_ftrl} will result in the trajectory $\mathcal{T}_2$ given by Table~\ref{tab:algorithm_behavior_otherwise} where convergence occurs by $$t_2' = \Bigg\lfloor \frac{2(1-p_{\min})-(1-p_{\max})}{p_{\max}-p_{\min}}+\frac{2}{M(p_{\max} - p_{\min})}\Bigg\rfloor+1.$$ 
\end{itemize}

\begin{table}[h]
    \centering
    \begin{tabular}{|c|c|c|}
    \hline
         Time & $w_p^{(t)}$  & $w_r^{(t)}$  \\\hline
          $t=1$ & $w_{p,1}^{(1)}$ & $w_{r,1}^{(1)}$\\
          $t=2$ & $w_{r,1}^{(1)}$ & $\min\{\alpha_r, w_{p,1}^{(1)}\}$\\
          $t\ge 3$ & $p_{\min}$ & $p_{\min}$\\\hline
    \end{tabular}
    \caption{Trajectory $\mathcal{T}_1$ of Algorithm~\ref{alg:modified_ftrl} with initial conditions $\mathcal{C}_1$.}
    \label{tab:algorithm_behavior_t_3}
\end{table}
\begin{table}[h]
    \centering
    \begin{tabular}{|c|c|c|}
    \hline
         Time & $w_p^{(t)}$  & $w_r^{(t)}$  \\\hline
          $t=1$ & $w_{p,1}^{(1)}$ & $w_{r,1}^{(1)}$\\
          $t=2$ & $w_{r,1}^{(1)}$ & $\min\{\alpha_r, w_{p,1}^{(1)}\}$\\
          $3\le t < t_2'$ & $p_{\max}$ & $p_{\min}$\\
          $t\ge t_2'$ & $p_{\min}$ & $p_{\min}$\\\hline
    \end{tabular}
    \caption{Trajectory $\mathcal{T}_2$ of Algorithm~\ref{alg:modified_ftrl} with initial conditions $\mathcal{C}_2$.}
    \label{tab:algorithm_behavior_otherwise}
\end{table}

Here, note that $p_{\min}$ is the smallest offer that agent $R$ has seen from agent $P$ up until time $t=3$. So, both $\mathcal{T}_1$ and $\mathcal{T}_2$ converge to this smallest offer. The intuition here is that the game $\mathcal{G}^1$ has a weakly dominant strategy for agent $R$ to accept any offer (since otherwise both agents get 0), so the influence of the regularizer is minimized with a large enough $M$ and it is optimal for agent $R$ to converge to this strategy. Following this, it is initially better for agent $P$ to make an offer that is accepted in all rounds, given by $p_{\max}$, but as agent $R$ continues to choose $p_{\min}$, then it becomes optimal for agent $P$ to give up one deal to strike a larger deal for themselves in every other round. 

We will now directly show the behavior, i.e., which strategies are chosen, of $\mathcal{T}_1$ and $\mathcal{T}_2$ at the base cases of $t=1$ to $t=3$ and then we will induct on $t$ to prove the rest of these two trajectories holds. 

\textbf{Behavior at $t=1$.}
At $t=1$, each agent plays their arbitrary starting parameter as a pure strategy by definition of Algorithm~\ref{alg:modified_ftrl}. That is, for both $\mathcal{T}_1$ and $\mathcal{T}_2$, $w_r^{(1)} = w_{r,1}^{(1)}$ and $w_p^{(1)} = w_{p,1}^{(1)}$.\\ 

\textbf{Behavior at $t=2$.}
First, consider agent $R$. The objective function for a possible maximizing strategy $s_R$ is $$U_R^{(2)}(s_R, \{w_{p,1}^{(1)}\}) - \frac{\|s_R - \alpha_r\|_1}{M}.$$ From Lemma~\ref{lem:maximizers}, the set of possible maximizers is $\{w_p^{(1)} = w_{p,1}^{(1)}, \alpha_r\}$. We can break the analysis up into two cases to conclude that, for both $\mathcal{T}_1$ and $\mathcal{T}_2$, $w_r^{(2)} = \min\{\alpha_r, w_{p,1}^{(1)}\}$.

\textbf{Case 1,} if 
 $\alpha_r \le w_{p,1}^{(1)}$, then, trivially, we have $w_r^{(2)} = \alpha_r$ since $s_R=\alpha_r$ maximizes $U_R^{(2)}(s_R, \{w_{p,1}^{(1)}\})$ when $\alpha_r \le w_{p,1}^{(1)}$ and minimizes $\frac{\|s_R - \alpha_r\|_1}{M}$.
 
 \textbf{Case 2,} if $w_{p,1}^{(1)} < \alpha_r$, then we can examine the objective value of each possible maximizing strategy. We have that $w_p^{(1)} = w_{p,1}^{(1)}$, so the objective value for $s_R = w_{p,1}^{(1)}$ is $$w_{p,1}^{(1)} - \frac{2}{M},$$ and the objective value for $s_R = \alpha_r$ is $0$ since an acceptance threshold of $\alpha_r$ is greater than an offer of $w_{p,1}^{(1)}$ in this case. Then, the objective value for $s_R = w_{p,1}^{(1)}$ is strictly greater than that of $\alpha_r$ when $$M > (\frac{2}{w_{p,1}^{(1)}})^2$$ which is satisfied by the setting of $M$ above. Thus, in either case, for both $\mathcal{T}_1$ and $\mathcal{T}_2$, we have $w_r^{(2)} = \min\{\alpha_r, w_{p,1}^{(1)}\}$.\\

Next, consider agent $P$. The objective function for a possible maximizing strategy $s_P$ is $$U_P^{(2)}(s_P, \{w_{r,1}^{(1)}\}) - \frac{\|s_P - \alpha_p\|_1}{M}.$$ From Lemma~\ref{lem:maximizers}, the set of possible maximizers is $\{w_r^{(1)} = w_{r,1}^{(1)}, \alpha_p\}$. We can again break the analysis up into two cases to conclude that, for both $\mathcal{T}_1$ and $\mathcal{T}_2$, $w_p^{(2)} = w_{r,1}^{(1)}$.

\textbf{Case 1,} if $\alpha_p=w_{r,1}^{(1)}$, then, trivially, we will have $w_p^{(2)} = w_{r,1}^{(1)}$ since $\alpha_p$ and $w_{r,1}^{(1)}$ are the only two possible maximziers.

\textbf{Case 2,} otherwise, if $\alpha_p \neq w_{r,1}^{(1)}$, the objective value for $s_P = w_{r,1}^{(1)}$ is $$1-w_{r,1}^{(1)} - \frac{2}{M},$$and the objective value at $s_P = \alpha_p$ is $$U_P^{(2)}(\alpha_p, \{w_{r,1}^{(1)}\}).$$ Then, the objective value for $s_P = w_{r,1}^{(1)}$ is strictly greater than that of $\alpha_p$ when $$M > \left(\frac{2}{1-w_{r,1}^{(1)} - U_P^{(2)}(\alpha_p, \{w_{r,1}^{(1)}\})}\right)^2.$$ Again, this is satisfied by the setting of $M$ above. Note that the denominator here is only 0 if $\alpha_p = w_{r,1}^{(1)}$, so this term will be well-defined in this case. Thus, in either case, for both $\mathcal{T}_1$ and $\mathcal{T}_2$, we have $w_p^{(2)} = w_{r,1}^{(1)}$.\\

\textbf{Behavior at $t=3$.} First, consider agent $R$. The objective function for a possible maximizing strategy $s_R$ is $$U_R^{(3)}(s_R,\{w_{p,1}^{(1)}, w_{r,1}^{(1)}\}) - \frac{\|s_R- \alpha_r\|_1}{M}.$$From Lemma~\ref{lem:maximizers}, the set of possible objective maximizers is $\{w_p^{(1)}= w_{p,1}^{(1)}, w_p^{(2)} = w_{r,1}^{(1)}, \alpha_r \}$. We can break the analysis up into two cases to conclude that, for both $\mathcal{T}_1$ and $\mathcal{T}_2$, $w_r^{(3)} = p_{\min}$.

\textbf{Case 1,} if $p_{\min} = \alpha_r$, then trivially we have $\alpha_r$ as the maximizer since $s_R = \alpha_r$ maximizes $U_R^{(3)}(s_P,\{w_{p,1}^{(1)}, w_{r,1}^{(1)}\})$ since $p_{\min} = \alpha_r$ implies $\alpha_r \le w_{p,1}^{(1)}$ and $\alpha_r \le w_{r,1}^{(1)}$ and, further, $s_R = \alpha_r$  minimizes $\frac{\|s_R- \alpha_r\|_1}{M}$. 

\textbf{Case 2, } if $p_{\min} < \alpha_r$, then $s_R = p_{\min}$ has an objective value of 
$$w_{p,1}^{(1)} + w_{r,1}^{(1)} - \frac{2}{M}.$$
Note that $p_{\min}$ is equal to $\min\{w_{p,1}^{(1)},w_{r,1}^{(1)}\}$, and the set of possible objective maximizers includes $\alpha_r$ and $\max\{w_{p,1}^{(1)},w_{r,1}^{(1)}\}$, so it suffices to compare the objective values of $p_{\min}$ to that of $\alpha_r$ and $\max\{w_{r,1}^{(1)},w_{p,1}^{(1)}\}$ directly.

First, the objective value of $\alpha_r$ is $$U_R^{(3)}(\alpha_r,\{w_{p,1}^{(1)},w_{r,1}^{(1)}\}).$$ Then, the objective value of $p_{\min}$ is strictly greater than this objective value when$$M > \left(\frac{2}{w_{p,1}^{(1)} + w_{r,1}^{(1)} - U_R^{(3)}(\alpha_r,\{w_{p,1}^{(1)},w_{r,1}^{(1)}\})}\right)^2,$$ and this is true for the setting $M$. Note again the denominator is only 0 when $\alpha_r = p_{\min}$ which is not the case here such that we know the term is well-defined.  
Next, if $\alpha_r = \max\{w_{p,1}^{(1)},w_{r,1}^{(1)}\}$, then they have the same objective value, so suppose $\max\{w_{r,1}^{(1)}, w_{p,1}^{(1)}\} \neq \alpha_r$. Then, the objective value of $\max\{w_{r,1}^{(1)},w_{p,1}^{(1)}\}$ is less than or equal to that of $p_{\min}$ since the higher acceptance threshold gets strictly less utility (unless $w_{p,1}^{(1)}=w_{r,1}^{(1)}$ in which case they get equal utility) and each strategy incurs a regularization penalty of $\frac{2}{M}$. Thus, in all cases, for both $\mathcal{T}_1$ and $\mathcal{T}_2$, agent $R$ will choose $w_r^{(3)}=p_{\min}$.\\

Next, consider agent $P$. The objective function for a possible maximizing strategy $s_P$ is $$U_P^{(3)}(s_P, \{w_{r,1}^{(1)}, \min\{\alpha_r,w_{p,1}^{(1)}\}\}) - \frac{\|s_P- \alpha_p\|_1}{M}.$$From Lemma~\ref{lem:maximizers}, the set of possible objective maximizers is $\{w_r^{(1)} =w_{r,1}^{(1)}, w_r^{(2)} = \min\{\alpha_r,w_{p,1}^{(1)}\},\alpha_p\}$. Note that $p_{\min}$ will be $w_{r,1}^{(1)}$ or  $\min\{\alpha_r, w_{p,1}^{(1)}\}$ and  $p_{\max}$ will be the other. 

At $t=3$, $\mathcal{T}_1$ and $\mathcal{T}_2$ exhibit different behavior for agent $P$, so we will prove the behavior of each trajectory separately.  

\textbf{Behavior of $\mathcal{T}_1$ at $t=3$ for agent $P$.}
Recall that we are in the trajectory $\mathcal{T}_1$ when $\mathcal{C}_1$ is true. So, we will go through the three possible conditions that make $\mathcal{C}_1$ true and show that $p_{\min}$ is the objective maximizer.

\textbf{Condition 1.} Suppose $p_{\min} = p_{\max}$, then it suffices to 
 show the objective value of $p_{\min}$ is strictly greater than that of $\alpha_p$. Suppose $p_{\min} \neq \alpha_p$ (since otherwise the objective values are trivially equal), then the objective value at $p_{\min}$ is $$2(1-p_{\min}) - \frac{2}{M},$$and the objective value at $\alpha_p$ is $$U_P^{(3)}(\alpha_p,\{w_{r,1}^{(1)},\min\{\alpha_r,w_{p,1}^{(1)}\}\})$$ such that the objective value at $p_{\min}$ is strictly greater when $$M > \left(\frac{2}{2(1-p_{\min})-U_P^{(3)}(\alpha_p, \{w_{r,1}^{(1)}, \min\{\alpha_r, w_{p,1}^{(1)}\}\})}\right)^2.$$ Note that $M$ satisfies this.

 \textbf{Condition 2.} Suppose $p_{\min} \neq p_{\max}$ and $ 2(1-p_{\max})<1-p_{\min}$. Here, the utility that the strategy $p_{\min}$ gets is $$U_P^{(3)}(p_{\min}, \{w_{r,1}^{(1)}, \min\{\alpha_r,w_{p,1}^{(1)}\}\}) = 1-p_{\min},$$ and the utility that the strategy $p_{\max}$ gets is $$U_P^{(3)}(p_{\max}, \{w_{r,1}^{(1)}, \min\{\alpha_r,w_{p,1}^{(1)}\}\}) = 2(1-p_{\max})$$ since each strategy is an offer and the lower offer gets one acceptance and the higher offer gets two acceptances. If $\alpha_p \neq p_{\min}$ and $\alpha_p \neq p_{\max}$, then, both strategies incur a regularization penalty of $\frac{2}{M}$ and since $1-p_{\min} > 2(1-p_{\max})$, it is clear that $p_{\min}$ has a strictly larger objective value than $p_{\max}$. Further, $p_{\min}$ has a larger objective value than $\alpha_p$ when $$M > \left(\frac{2}{(1-p_{\min})-U_P^{(3)}(\alpha_p, \{w_{r,1}^{(1)}, \min\{\alpha_r, w_{p,1}^{(1)}\}\})}\right)^2$$ which is satisfied by the setting of $M$. To continue, if $\alpha_p = p_{\min}$, then $p_{\min}$ is trivially still the objective maximizing strategy since it maximizes $U_P^{(3)}$ and minimizes the regularization term. Finally, if $\alpha_p = p_{\max}$, then the setting of $M$ above still holds to make $p_{\min}$ the objective maximizing strategy since $$1-p_{\min} > 2(1-p_{\max}) = U_P^{(3)}(\alpha_p, \{w_{r,1}^{(1)}, \min\{\alpha_r, w_{p,1}^{(1)}\}\}).$$

 \textbf{Condition 3.} Suppose $p_{\min} \neq p_{\max}$ and $\alpha_p = p_{\min}$ and $2(1-p_{\max}) = 1-p_{\min}$. Since $\alpha_p = p_{\min}$, it remains to show that the objective value of $p_{\min}$ is strictly larger than that of $p_{\max}$. Since $2(1-p_{\max}) = 1-p_{\min}$, the utility gained from each strategy is the same. However, since $\alpha_p = p_{\min}$ and $p_{\max} \neq p_{\min}$, then $p_{\max}$ incurs a regularization penalty of $\frac{2}{M}$ while $p_{\min}$ incurs no regularization penalty. So, $p_{\min}$ must have a strictly larger objective value that $p_{\max}$.

 So, in all conditions, $w_p^{(3)} = p_{\min}$ as desired. Further, since we showed above $w_r^{(3)} = p_{\min}$ as well, we can conclude Algorithm~\ref{alg:modified_ftrl} is in a pure NE in $\mathcal{G}^1$ and has converged at $t_1' = 3$. The induction proof which shows why Algorithm~\ref{alg:modified_ftrl} stays at any pure NE once it reaches one remains to be shown in the final section of this proof.

\textbf{Behavior of $\mathcal{T}_2$ at $t=3$ for agent $P$.}
Recall that we are in the trajectory $\mathcal{T}_2$ when $\mathcal{C}_2$ is true. So, we will go through the two possible conditions that make $\mathcal{C}_2$ true and show that $p_{\max}$ is the objective maximizer.

\textbf{Condition 1.} Suppose $p_{\min} \neq p_{\max}$ and $2(1-p_{\max}) > 1-p_{\min}$. As in the case of $\mathcal{T}_1$ above, the utility that the strategy $p_{\min}$ gets is $$U_P^{(3)}(p_{\min}, \{w_{r,1}^{(1)}, \min\{\alpha_r,w_{p,1}^{(1)}\}\}) = 1-p_{\min},$$ and the utility that the strategy $p_{\max}$ gets is $$U_P^{(3)}(p_{\max}, \{w_{r,1}^{(1)}, \min\{\alpha_r,w_{p,1}^{(1)}\}\}) = 2(1-p_{\max}).$$ If $\alpha_p \neq p_{\min}$ and $\alpha_p \neq p_{\max}$, then both strategies are also both incur a regularization penalty of $\frac{2}{M}$ and $p_{\max}$ clearly has a larger objective value given our assumptions. Further, $p_{\max}$ has a larger objective value compared to that of $\alpha_p$ when $$M > \left(\frac{2}{2(1-p_{\max})-U_P^{(3)}(\alpha_p, \{w_{r,1}^{(1)}, \min\{\alpha_r, w_{p,1}^{(1)}\}\})}\right)^2$$ which is true given the setting of $M$. Note that the denominator is only 0 when $\alpha_p = p_{\max}$ or $2(1-p_{\max}) = U_P^{(3)}(\alpha_p, \{w_{r,1}^{(1)}, \min\{\alpha_r, w_{p,1}^{(1)}\}\})$ and both of these cases are not possible given our assumptions. If $\alpha_p = p_{\max}$, then $p_{\max}$ is trivially the objective maximizer since it still maximizes the utility and it now minimizes the regularization function as well. Finally, if $\alpha_p = p_{\min}$, then the setting of $M$ above still holds to make $p_{\max}$ the objective maximizing strategy since $$2(1-p_{\max}) > 1-p_{\min} = U_P^{(3)}(\alpha_p, \{w_{r,1}^{(1)}, \min\{\alpha_r, w_{p,1}^{(1)}\}\}).$$

\textbf{Condition 2.} Suppose $p_{\min} \neq p_{\max}$ and $2(1-p_{\max}) = 1-p_{\min}$ and $\alpha_p \neq p_{\min}$. Here, if $\alpha_p \neq p_{\max}$, then the objective value of $p_{\max}$ is strictly larger than that of $\alpha_p$ by the same setting of $M$ above. Further, $p_{\max}$ and $p_{\min}$ have the same objective value, so, by Algorithm~\ref{alg:modified_ftrl}'s definition, $p_{\max}$ is chosen to break the objective maximizing tie. Otherwise, if $\alpha_p = p_{\max}$, then $p_{\max}$ is the objective maximizer because it maximizes the utility and minimizes the regularization penalty. 

So, in each condition, $w_p^{(3)} = p_{\max}$ as desired. We move to the next section to induct on $t$ to show that this behavior will continue in $\mathcal{T}_2$ until time $t_2'$ defined above.

\textbf{Induction Proof 1: Behavior of $\mathcal{T}_2$ during $3 < t < t_2'$}\\
Let $t$ be a time step such that $3 < t < t_2'$. Suppose each agent behaves according to Table~\ref{tab:algorithm_behavior_otherwise} through time step $3$ and then for all $3 <\tau \le t-1$ suppose we have $$w_p^{(\tau)} = p_{\max}$$ and $$w_r^{(\tau)} = p_{\min}.$$ 

We will now show that at the next time step $t$, the maximizer for agent $P$ will still be $p_{\max}$ and the maximizer for agent $R$ will still be $p_{\min}$. 

Consider agent $R$. At time $t$, the next maximizer that agent $R$ chooses is $$w_r^{(t)} = \argmax_{w \in \Delta(\mathcal{S})} U_R^{(t-1)}(w, \{w_{p,1}^{(1)},w_{r,1}^{(1)}, p_{\max}^{(3:t-1)}\}) - \frac{\|w-\alpha_r\|_1}{M}.$$
 From the base cases, it is true that $p_{\min}$ is the objective maximizer at time $\tau=3$ and, further, no other acceptance threshold strategy gets strictly more utility from time $\tau=4$ to time $\tau=t-1$ since any acceptance threshold less than or equal to the offer of $p_{\max}$ gets utility of $p_{\max}$. Thus, it will be the case that $p_{\min}$ will continue to be the objective maximizer and we can conclude $w_r^{(t)} = p_{\min}$.

Next, consider agent $P$. At time $t$, the next maximizer that agent $P$ chooses is $$w_p^{(t)} = \argmax_{w \in \Delta(\mathcal{S})}U_P^{(t-1)}(w, \{w_{r,1}^{(1)},\min\{\alpha_r,w_{p,1}^{(1)}\}, p_{\min}^{(3:t-1)}\})  - \frac{\|w-\alpha_p\|_1}{M}.$$
From the base cases, it is true that $p_{\max}$ is the objective maximizer at time $\tau=3$ and, further, $p_{\min}$ maximizes the utility functions from $\tau=4$ to $\tau=t-1$. Thus, it suffices to compare the objective value at time $t$ between $p_{\max}$ and $p_{\min}$ and the same arguments as in the base case of $t=3$ hold for why $\alpha_p$ is not the objective maximizer. The objective value for $p_{\max}$ is $$(t-1)(1-p_{\max}) - \frac{\|p_{\max}-\alpha_p\|_1}{M}$$ and for $p_{\min}$ is $$(t-2)(1-p_{\min}) - \frac{\|p_{\max}-\alpha_p\|_1}{M}.$$ If $\alpha_p \neq p_{\min}$ and $\alpha_p \neq p_{\max}$, then the objective value for $p_{\max}$ is greater than or equal to that of $p_{\min}$ when $$t \le \frac{2(1-p_{\min})-(1-p_{\max})}{p_{\max}-p_{\min}}.$$ Next, if $\alpha_p = p_{\max}$, then the objective value for $p_{\max}$ is greater than or equal to that of $p_{\min}$ when $$t \le \frac{2(1-p_{\min})-(1-p_{\max})}{p_{\max}-p_{\min}} + \frac{2}{M(p_{\max}-p_{\min})}.$$Finally, if $\alpha_p = p_{\min}$, then the objective value for $p_{\max}$ is greater than or equal to that of $p_{\min}$ when $$t \le \frac{2(1-p_{\min})-(1-p_{\max})}{p_{\max}-p_{\min}} - \frac{2}{M(p_{\max}-p_{\min})}.$$From our definition of $t_2'$ it is true that $t$ either satisfies the expression and we can conclude that $w_p^{(t)} = p_{\max}$ and the inductive step holds or it is the case that $t$ exceeds the upper bound and we can instead conclude $w_p^{(t)} = p_{\min}$ and that convergence has occurred earlier than $t_2'$.

From above, note that at time $t=t_2'$, it is now the case that$$t_2' > \frac{2(1-p_{\min})-(1-p_{\max})}{p_{\max}-p_{\min}} + \frac{2}{M(p_{\max}-p_{\min})},$$ which means $t_2'$ exceeds each of the three bounds above and we can conclude that $w_p^{(t_2')} = p_{\min}$ and that convergence occurs, at the latest, at $t_2'$ in trajectory $\mathcal{T}_2$.

\textbf{Induction Proof 2: Behavior after $t_1'$ for $\mathcal{T}_1$ and $t_2'$ for $\mathcal{T}_2$}\\
We know that at time $t'=t_1'$ for $\mathcal{T}_1$ and time $t'=t_2'$ for $\mathcal{T}_2$, we will reach a strategy profile $(w_p^{t'},w_r^{t'})$ that is in Nash equilibrium where $w_p^{t'} = w_r^{t'} = p_{\min}$. That is, for each agent $i \in \{P,R\}$,

$$w_i^{t'} = \argmax_{w \in \Delta(S)} U_i^{(t')}(w, w_{j}^{1:t'-1}) - \frac{\|w-\alpha_i\|_1}{M} = p_{\min}.$$
The next step of the optimization includes the utility agent $i$ gets at time $t'$ and can be written as 
$$w_i^{t'+1} = \argmax_{w \in \Delta(S)} U_i^{(t')}(w, w_{j}^{1:t'-1})  - \frac{\|w-\alpha_i\|_1}{M} + u_i(w, w_{j}^{t'}).$$
Since $w_i^{t'}$ maximizes the first expression by definition and the second expression by the fact that $(w_p^{t'}, w_r^{t'})$ is in NE, we can conclude that $w_i^{t'}$ maximizes the entire objective for all future time steps.\\

Thus, the behavior of Algorithm~\ref{alg:modified_ftrl} follows Table~\ref{tab:algorithm_behavior_t_3} or Table~\ref{tab:algorithm_behavior_otherwise} when $M$ satisfies a set of inequalities determined by the values of $\{w_{p,1}^{(1)},w_{r,1}^{(1)},\alpha_p,\alpha_r\}$. Therefore, there exists an $M >0 $ such that agents $P$ and $R$ learning a strategy from Algorithm~\ref{alg:modified_ftrl} converge to a NE at some time step $t'\ge 3$. Further, the value of the NE is $\min\{\alpha_r, w_{p,1}^{(1)},w_{r,1}^{(1)}\}$ and the final strategy profile is also a NE in the game $\mathcal{G}^1$. 
\end{proof}

\begin{customthrm}{5}
\label{thrm:n=2_convergence}
    Suppose the following initial conditions hold:
    \begin{align*}
        1-w_{r,1}^{(1)} &\ge \delta w_{r,2}^{(1)}, & \label{thrm_n=2:asmp_1}\tag{1}\\
        w_{p,1}^{(1)} &> \delta(1-w_{p,2}^{(1)}), \text{ and } & \label{thrm_n=2:asmp_2}\tag{2}\\
        \alpha_{i,1} &> w_{j,1}^{(1)}. &\label{thrm_n=2:asmp_3}\tag{3}
    \end{align*}
    Then, there exists a time $t'$ such that Algorithm~\ref{alg:modified_ftrl}, parameterized by a discretization constant $D>\frac{1}{1-\delta}$, learning rate $M > 2D$, and $p=1$, converges to a NE in $\mathcal{G}^{(2)}$.
\end{customthrm}

To begin, we make use of the following lemmas.
\begin{customlem}{3}
\label{lem:utility_maximizing}
    Suppose after time $t$, the agents play the strategy profile $(w_p^{(t)}, w_r^{(t)})$. If agent $i$ for $i \in \{P, R\}$ maximizes utility with $w_i^{(t)}$ with respect to $w_{j}^{(t)}$, then agent $i$ will play $w_i^{(t+1)} = w_{i}^{(t)}$ in the next time step.
\end{customlem}
\begin{proof}
    By the update step of Algorithm~\ref{alg:modified_ftrl}, it is the case that $$w_i^{(t)} = \argmax_{w \in \Delta(\mathcal{S}_{i})} \sum_{\tau=1}^{t-1}u_i(w, w_{j}^\tau) - \frac{\|w-\alpha_i\|_1}{M}.$$
    So, if it is also the case that $$w_i^{(t)} = \argmax_{w \in \Delta(\mathcal{S}_{i})} u_i(w, w_{j}^{(t)}), $$ then it follows that $$w_i^{(t+1)} = \argmax_{w \in \Delta(\mathcal{S}_{i})} \sum_{\tau=1}^{t}u_i(w, w_{j}^\tau) - \frac{\|w-\alpha_i\|_1}{M} = w_i^{(t)}.$$
\end{proof}

\begin{customlem}{4}
\label{lem:non_utility_maximizing}
    Suppose after time $t$, the agents play the strategy profile $(w_p^{(t)}, w_r^{(t)})$. If agent $i$ for $i \in \{P, R\}$ does not maximize utility with $w_i^{(t)}$ with respect to $w_{j}^{(t)}$, then there exists a time step $t'> t$ where agent $i$ will play a strategy which gets strictly more utility than $w_i^{(t)}$ with respect to $w_{j}^{(t)}$.
\end{customlem}
\begin{proof}
    To begin, let $\mathcal{W}_i(w_i, w_{j})$ be the set of strategies for agent $i$ that gets strictly more utility than the strategy $w_i$ with respect to $w_{j}$. This set is defined as follows:
    $$\mathcal{W}_i(w_i, w_{j}) = \{w \in \Delta(\mathcal{S}_i)| u_i(w, w_{j}) > u_i(w_i, w_{j}) \}.$$
    By the update step of Algorithm~\ref{alg:modified_ftrl}, it is the case that $$w_i^{(t)} = \argmax_{w \in \Delta(\mathcal{S}_{i})} \sum_{\tau=1}^{t-1}u_i(w, w_{j}^\tau) - \frac{\|w-\alpha_i\|_1}{M}.$$
    However, it is now the case that $$u_i(w_i^{(t)}, w_{j}^{(t)}) < \argmax_{w \in \Delta(\mathcal{S}_{i})} u_i(w, w_{j}^{(t)}).$$ 
    We assume agent $j$ is currently utility-maximizing with $w_{j}^{(t)}$ with respect to $w_i^{(t)}$ such that, by lemma~\ref{lem:utility_maximizing}, agent $j$ will continue to choose $w_{j}^{(t)}$ at subsequent time steps. 

    Let $\mathcal{O}_i^{(t)}(w)$ be the objective value of strategy $w$ for agent $i$ up until time $t$. That is, $$\mathcal{O}_i^{(t)}(w) :=  \sum_{\tau=1}^{t-1}u_i(w, w_{j}^\tau) - \frac{\|w-\alpha_i\|_1}{M}.$$

    Then, at time $t$, we have a set of $|\mathcal{W}_i(w_i^{(t)}, w_{j}^{(t)})|$ affine functions mapping the number of time steps since (and including) $t$ to the objective value of a strategy $w$. Each function has a slope given by the utility of $w \in \mathcal{W}_i(w_i^{(t)}, w_{j}^{(t)})$ with respect to $w_{j}^{(t)}$ and a corresponding constant of $\mathcal{O}_i^{(t)}(w)$. Formally, for each $w \in  \mathcal{W}_i(w_{i}^{(t)}, w_{j}^{(t)})$, let $f_{w}^{(t)}: \mathbb{N} \to \mathbb{R}$ be defined as $$f_{w}^{(t)}(n) = u_i(w, w_{j}^{(t)}) \cdot n + \mathcal{O}_i^{(t)}(w).$$

    Then, the first $w \in \mathcal{W}_i(w_i^{(t)}, w_{j}^{(t)})$ such that for some $n >0$, $f_w^{(t)}(n) > f_{w_i^{(t)}}^{(t)}(n)$ (or the largest valued $w$ if there are ties) will be chosen at time step $t' = t+n$. 
\end{proof}

\begin{customlem}{5}
\label{lem:following_behavior_P2R1}
    Suppose each agent $P$ and $R$ is using Algorithm~\ref{alg:modified_ftrl} parameterized by $D >1,M > 2D,$ and $ p=1$. Suppose after updating at time $t$ we get a strategy profile $(w_p^{(t)}, w_r^{(t)})$ where \begin{align*}
        w_{r,1}^{(t)} &= w_{p,1}^{(t)} = c,&\label{asmp:lem4_asmp1}\tag{1}\\
         1-c&< \delta\cdot w_{r,2}^{(t)}, &\label{asmp:lem4_asmp2}\tag{2}\\
        c &> \delta\cdot (1-w_{p,2}^{(t)}),&\label{asmp:lem4_asmp3}\tag{3}\\
        \alpha_{r,1} &> w_{p,1}^{(t)},&\label{asmp:lem4_asmp4}\tag{4}\\
        \alpha_{p,1} &> w_{r,1}^{(t)}.&\label{asmp:lem4_asmp5}\tag{5}\\
    \end{align*}
    Suppose further, that $w_{p,1}^{(t)} = w_{r,1}^{(t)}= c$ is the lowest acceptance threshold and offer in the first round for the responder or the proposer, respectively, and that $w_{p,2}^{(t)}$ is the lowest offer $P$ has seen in round two, i.e., $w_{p,2}^{(t)} = \min\{w_{r,2}^{(1)},\ldots,w_{r,2}^{(t)}\}$.  Then, there exists a time step $t' > t$ where the agents reach a NE.
\end{customlem}
\begin{proof}
    We will start by showing that there exists a time $t_1' > t$ where $P$ begins to lower their first round offer to get a preferred second round deal. We will then show that $R$ will lower their first round acceptance threshold and continue to follow $P$ until $R$ would prefer to make a second round deal as well at some time $t_2' > t_1'$. Finally, we will show that the agents end with either a second round deal in NE or $P$ concedes to raise their first round offer slightly to also be in a NE.

    At time $t+1$ for agent $R$, by assumption~\ref{asmp:lem4_asmp3}, the current objective-maximizing strategy $w_r^{(t)}$ is also the utility maximizer at time $t$. So, by lemma~\ref{lem:utility_maximizing}, we can conclude $w_r^{(t+1)} = w_r^{(t)}$.

    Next, at time $t+1$ for agent $P$, by assumption~\ref{asmp:lem4_asmp2}, the proposer's utility-maximizing strategy given $w_r^{(t)}$ is to choose a first round offer of $c-\frac{1}{D}$ and a second round acceptance threshold  less than or equal to $w_{r,2}^{(t)}$ to get a second round deal of $\delta \cdot w_{r,2}^{(t)}$. Since $w_{p,2}^{(t_1')} = \min\{w_{r,2}^{(1)},\ldots,w_{r,2}^{(t)}\}$, then $w = (c-\frac{1}{D}, w_{p,2}^{(t)})$ is the strategy that gets the most utility, given $w_r^{(t)}$ and the entire history of previous strategy profiles. Further, $w$ gets strictly more utility than $w_p^{(t)}$  by assumption~\ref{asmp:lem4_asmp2}, so by lemma~\ref{lem:non_utility_maximizing}, there exists a time step $t_1' > t$ where $w_p^{(t_1')} = (c-\frac{1}{D}, w_{p,2}^{(t)})$. So, at time $t_1' > t$, we are in the strategy profile $$(w_p^{(t_1')}, w_r^{(t_1')}) = \left((c-\frac{1}{D}, w_{p,2}^{(t)}),(c, w_{r,2}^{(t)})\right).$$

    Next, at time $t_1'+1$, agent $P$'s strategy of $w_p^{(t_1')}$ is  utility-maximizing with respect to $w_{r}^{(t_1')}$. So, by lemma~\ref{lem:utility_maximizing}, $w_p^{(t_1'+1)} = w_p^{(t_1')}$. Next, agent $R$, given $w_p^{(t_1')}$, will either prefer to make a first or second round deal. If $R$ prefers a second round deal, then by lemma~\ref{lem:non_utility_maximizing}, there exists a time $t_2'\ge t_1'+1$ where the agents are in the strategy profile
    $$(w_p^{(t_2')}, w_r^{(t_2')}) = \left((c-\frac{1}{D}, w_{p,2}^{(t)}),(c, w_{p,2}^{(t)})\right).$$ 
    This strategy profile is either in NE if $1-c < \delta w_{p,2}^{(t)}$ or, by lemma~\ref{lem:non_utility_maximizing}, there exists a time $t_3' \ge t_2'+1$ where $P$ raises their first round offer back to $c$ without changing their second round acceptance threshold, and then assumption~\ref{asmp:lem4_asmp3} shows that the strategy profile would be in NE. 
    
    Otherwise, if $R$ prefers a first round deal at time $t_1'+1$, then by lemma~\ref{lem:non_utility_maximizing}, there exists a time step $\tau_1 \ge t_1'+1$ where $R$ will choose a strategy of $w_r^{(\tau_1)}= (c-\frac{1}{D}, 1)$, since $c-\frac{1}{D}$ is the lowest first round offer so far, so agent $R$ is making all first round deals and the second round strategy parameter is set to 1 by the tie-breaking mechanism. Then, if $\delta \le (1-(c-\frac{1}{D}))$ such that $P$ also prefers a first round deal, then the strategy profile is in NE by definition at time $\tau_1 \ge t_1'+1$. Otherwise, if $\delta >  (1-(c-\frac{1}{D}))$, then $P$ prefers a second round deal, and by lemma~\ref{lem:non_utility_maximizing}, there exists a time $\tau_2>\tau_1$ such that $P$ switches to the utility-maximizing strategy of $w_p^{(\tau_2)}= (c-\frac{2}{D}, w_{p,2}^{(t)})$. Note that the second round acceptance threshold strategy does not change now since $w_{p,2}^{(t_1')}$ is still less than or equal to all the previous second round offers from $R$. By lemma~\ref{lem:utility_maximizing}, agent $R$ has not changed strategies, so at time $\tau_2$, we are in the strategy profile
    $$(w_p^{(\tau_2)}, w_r^{(\tau_2)}) = \left((c-\frac{2}{D}, w_{p,2}^{(t)}), (c-\frac{1}{D}, 1)\right).$$

    This behavior continues until $c-\frac{k}{D} \le \delta (1-w_{p,2}^{(t_1')})$ or $1-(c-\frac{k}{D}) \ge \delta$ for some $k \ge 2$. There exists such a $k$ since $w_{p,2}^{(t_1')} > 0$ and $c-\frac{k}{D} < c-\frac{k-1}{D}$. Let $\tau_k$ be the time step such that $c-\frac{k}{D} \le \delta (1-w_{p,2}^{(t_1')})$ or $1-(c-\frac{k}{D}) \ge \delta$. 
    
    If it is first the case that $1-(c-\frac{k}{D}) \ge \delta$, then this corresponds to $P$ no longer lowering their first round offer, and the agents will be in the strategy profile $$(w_p^{(\tau_k)}, w_r^{(\tau_k)}) = \left((c-\frac{k}{D}, w_{p,2}^{(t)}),(c-\frac{k}{D}, 1)\right)$$ which will be in NE.  
    
    Otherwise, if it is first the case that $c-\frac{k}{D} \le \delta (1-w_{p,2}^{(t_1')})$, then by lemma~\ref{lem:non_utility_maximizing}, there exists a time $t_2' \ge \tau_k$ such that $R$ switches to the strategy $(c-\frac{k-1}{D}, w_{p,2}^{(t)})$. Now, the agents are in the strategy profile $$(w_p^{(t_2')}, w_r^{(t_2')}) = \left((c-\frac{k}{D},w_{p,2}^{(t_1')} ),(c-\frac{k-1}{D}, w_{p,2}^{(t)}) \right).$$

    If $P$ also prefers the second round such that $\delta w_{p,2}^{(t)} > 1-(c-\frac{k-1}{D})$, then this strategy profile is in NE. If it is otherwise the case that $\delta w_{p,2}^{(t)} \le 1-(c-\frac{k-1}{D})$, then there exists a time $t_3' > t_2'$ where $P$ switches to the strategy profile $(c-\frac{k-1}{D}, w_{p,2}^{(t)})$. Since $R$ would only choose a first round acceptance threshold of $c-\frac{k-1}{D}$ if it was the case that $c-\frac{k-1}{D} > \delta (1-w_{p,2}^{(t_1')})$, then we can conclude their final following strategy profile is in NE:
    $$(w_p^{(t_3')}, w_r^{(t_3')}) =\left((c-\frac{k-1}{D},w_{p,2}^{(t_1')} ),(c-\frac{k-1}{D}, w_{p,2}^{(t)}) \right).$$

\end{proof}
Now, we are ready to prove Theorem~\ref{thrm:n=2_convergence}.\\

\begin{proof}
    At a high level, this proof will show that each agent will always switch to a utility-maximizing strategy, given their opponent's current strategy, eventually agents will agree in the first or second round (equal offer and acceptance threshold), and that agents will either be in a NE here or make a series of adjustments to reach a NE, depending on the initial conditions. By definition of Algorithm~\ref{alg:modified_ftrl}, agent $P$ starts with an arbitrary initial strategy $w_p^{(1)} = (w_{p,1}^{(1)}, w_{p,2}^{(1)})$ and regularizer reference point $\alpha_p = (\alpha_{p,1},\alpha_{p,2})$. Similarly, agent $R$ starts with an arbitrary initial strategy $w_r^{(1)} = (w_{r,1}^{(1)}, w_{r,2}^{(1)})$ and regularizer reference point $\alpha_r = (\alpha_{r,1},\alpha_{r,2})$. We will refer to $w_{p}^{(1)}, w_r^{(1)}, \alpha_p, \alpha_r$ as the \textit{initial conditions} of Algorithm~\ref{alg:modified_ftrl}. For ease of analysis, we assume none of the initial conditions are equal to $0$ or $1$ and we further assume $\alpha_p, \alpha_r$ are distinct from $w_p^{(1)}$ and $w_r^{(1)}$, respectively, in both dimensions. 

    By lemma~\ref{lem:ref_point_influence} and the setting of $M > 2D$, the strategy that is utility-maximizing in hindsight at time $t$ will always be chosen such that $\alpha_i$ will only be chosen if it is utility-maximizing.
    
    We can now directly show which strategies are utility-maximizing at each time step to show that there exists a time step $t'$ where the agents choose strategies that are in NE and the algorithm remains there.
    
    At time $t=2$, each agent $i$ will choose a strategy that maximizes their utility given $w_{j}^{(1)}$. Assumptions~\ref{thrm_n=2:asmp_1} and~\ref{thrm_n=2:asmp_2} indicate that the agent $P$ and $R$ maximize utility at time step $t=1$ with a first round deal, respectively. So, the utility-maximizing strategy for agent $R$ is the strategy  $$w_r^{(2)} =
        (w_{p,1}^{(1)},1),$$ since $\alpha_{r,1} > w_{p,1}^{(1)}$ by assumption~\ref{thrm_n=2:asmp_3}, and for agent $P$ is the strategy $$w_p^{(2)} = (w_{r,1}^{(1)},1).$$Note that any second round strategy parameter produces equal utility given each agent's first round strategy parameter, but our tie-breaking mechanism of choosing the largest valued strategy parameter (a second round strategy of 1 in this case) leads to the choice of the strategies above.  
    
    At time $t=3$, we can first see it is the case that $R$ prefers to make a first round deal again due to the choice of $w_{p}^{(2)}$ which gives agent $R$ 0 utility in the second round. So, the strategy agent $R$ chooses is $$w_r^{(3)} =
        (\min\{w_{p,1}^{(1)}, w_{p,1}^{(2)}\},1).$$ 
    
    For agent $P$, they must maximize the following sum of utilities with a strategy $w=(w_{p,1}, w_{p,2})$, i.e., $u_p(w,w_r^{(1)})+u_p(w,w_r^{(2)})$,:
    \begin{align*}
         \begin{cases}
        2(1-w_{p,1}) & w_{p,1} \ge \max\{w_{r,1}^{(1)}, w_{r,1}^{(2)}\}\\
        (1-w_{p,1}) + \delta \cdot w_{r,2}^{(\argmax\{w_{r,1}^{(1)}, w_{r,1}^{(2)}\})} & \min\{w_{r,1}^{(1)}, w_{r,1}^{(2)}\} \le w_{p,1}, \\
        &w_{p,1} < \max\{w_{r,1}^{(1)}, w_{r,1}^{(2)}\},\\
        &w_{p,2} \le w_{r,2}^{(\argmax\{w_{r,1}^{(1)}, w_{r,1}^{(2)}\})}\\
        \delta\cdot(w_{r,2}^{(1)} + w_{r,2}^{(2)}) & w_{p,1} < \min\{w_{r,1}^{(1)}, w_{r,1}^{(2)}\}, \\&w_{p,2} \le \min\{w_{r,2}^{(1)}, w_{r,2}^{(2)}\}\\
        0 & \text{otherwise}
    \end{cases}
    \end{align*}
    
    Since this is a piecewise linear function, the possible utility-maximizing strategy values for the proposer are at the endpoints of each piece (or the max possible value if many points get the same utility by the tie-breaking mechanism, or the reference point if it is utility-maximizing). These strategies are given with an index below and we will proceed by considering the case where each strategy is the maximizer at $t=3$ separately. Note that since $\alpha_{p,1} > w_{r,1}^{(1)}$ by assumption~\ref{thrm_n=2:asmp_3} and that $\alpha_p$ has distinct values from the initial points, it will always be the case that one of the strategies below will get strictly more utility than $\alpha_p$. \\
    \begin{table}[h]
        \centering
        \begin{tabular}{|c|c|c|}
    \hline
         Strategy & $w_{p,1}$ &$w_{p,2}$ \\\hline
          $w_1$ & $\max\{w_{r,1}^{(1)}, w_{r,1}^{(2)}\}$ & 1 \\ \hline
          $w_2$ & $\min\{w_{r,1}^{(1)}, w_{r,1}^{(2)}\} $ & $ w_{r,2}^{(\argmax\{w_{r,1}^{(1)}, w_{r,1}^{(2)}\})}$\\\hline
          $w_3$ & $\min\{w_{r,1}^{(1)},w_{r,1}^{(2)}\}-\frac{1}{D} $ & $\min\{w_{r,2}^{(1)}, w_{r,2}^{(2)}\} $ \\\hline
    \end{tabular}
        \caption{Possible maximizers at $t=3$ for $P$}
        \label{tab:case_1_maximizers}
    \end{table}

    Suppose strategy $w_1$ is utility-maximizing at $t=3$. Then, agent $R$ continues to maximize utility with the strategy $w_r^{(3)}$, so by lemma~\ref{lem:utility_maximizing}, it is the case that $w_r^{(t)} = w_r^{(3)}$ for all $t >3$ where $P$ chooses $w_1$. However, $w_1$ is not utility-maximizing for $P$, given $w_{r}^{(3)}$. So, by lemma~\ref{lem:non_utility_maximizing}, there exists a time $t_1' > 3$ where either $w_2$ or $w_3$ is chosen instead since these are the two strategies that get strictly more utility than $w_1$, given $w_r^{(3)}$. Therefore, at some time $t_1' > 3$, the agents will either have the strategy profile $(w_p^{(t')}, w_r^{(t')}) = (w_2, w_{r}^{(3)})$ or $(w_p^{(t_1')}, w_r^{(t_1')}) = (w_3, w_{r}^{(3)})$.

    Next, suppose strategy $w_2$ is utility-maximizing, either at time $t=3$ or $t=t_1'$ from above. Then, we are in the case where the first round strategies are equal to $\min\{w_{r,1}^{(1)}, w_{p,1}^{(1)}\}$. If both agents also get more utility in the first round than in the second round, given their opponent's strategy, then we can conclude that we are in NE by definition. Otherwise, at least one agent would prefer to make a deal in the second round. 
    \begin{itemize}
        \item Suppose agent $R$ prefers a second round deal. If $w_{r,1}^{(1)}<w_{p,1}^{(1)}$, then $w_2 = (w_{r,1}^{(1)}, 1)$ and $R$ would always prefer getting non-zero utility in the first round, so we can assume $w_{p,1}^{(1)} < w_{r,1}^{(1)}$ for the rest of this case. Then, by lemma~\ref{lem:non_utility_maximizing}, there exists a time step $t_2' > t$ where $R$ will switch to the higher first round acceptance threshold of $w_{r,1}^{(1)}$ and a second round offer of $w_{p,2}^{(1)}$ or $w_{r,2}^{(1)}$, depending on which is utility-maximizing in hindsight. Either $R$ will eventually switch to a strategy of $(w_{r,1}^{(1)}, w_{r,2}^{(1)})$ to utility-maximize with respect to $w_2$ or $P$ will switch a strategy of $(w_{r,1}^{(1)}, 1)$ if they prefer a round 1 deal. In the first case, $P$ will still switch to a strategy of $(w_{r,1}^{(1)}, 1)$ since they will prefer a round 1 deal given $R$'s strategy of $(w_{r,1}^{(1)}, w_{r,2}^{(1)})$ by assumption~\ref{thrm_n=2:asmp_1}. Whenever $P$ switches to $(w_{r,1}^{(1)}, 1)$, there is agreement in the first round that $P$ prefers and $R$ will also prefer this outcome to getting 0 utility in the second round and we can conclude we are in NE.
        \item If, instead, agent $P$ initially prefers a second round deal given the $R$ strategy  of $(\min\{w_{r,1}^{(1)}, w_{p,1}^{(1)}\}, 1)$, then, by lemma~\ref{lem:non_utility_maximizing} there exists a time $t_2' > t$ where they switch from strategy $w_2$ to strategy $w_3$ and we move to this case now.
        \item It's possible that both agents prefer the second round and switch strategies at the same time, but $R$ will still prefer round 2 to round 1 since $P$ will have lowered their first round offer and possible lowered their second round acceptance threshold, so the same behavior as in that case above still follows.
    \end{itemize}
    
    Finally, suppose strategy $w_3$ is chosen either at time $t=3, t_1',$ or $t_2'$ from above. By choosing $w_3$, $P$ must prefer the second round deal to making any first round deals, so it is now either the case that $R$ prefers the first or second round.
    
    \begin{itemize}
        \item If it is the case that $$\min\{w_{r,1}^{(1)}, w_{r,1}^{(2)}\} - \frac{1}{D} > \delta (1- \min\{w_{r,2}^{(1)}, w_{r,2}^{(2)}\}),$$
    then, $R$ should choose $(\min\{w_{r,1}^{(1)},w_{r,1}^{(2)}\}-\frac{1}{D}, 1)$. This is because a first round acceptance threshold of anything less than or equal to $\min\{w_{r,1}^{(1)},w_{r,1}^{(2)}\}$ is utility-maximizing in all the previous time steps since the second round acceptance threshold for $P$ was 1 in every time step except $t=1$, but there assumption~\ref{thrm_n=2:asmp_2} shows that a first round acceptance is preferable. Then, the conditions for lemma~\ref{lem:following_behavior_P2R1} are satisfied and we can conclude that this case will end in NE.
    \item Otherwise, if it is the case that $$\min\{w_{r,1}^{(1)}, w_{r,1}^{(2)}\} - \frac{1}{D} \le \delta (1- \min\{w_{r,2}^{(1)}, w_{r,2}^{(2)}\}),$$ then, $R$ should choose $(\min\{w_{r,1}^{(1)},w_{r,1}^{(2)}\}, \min\{w_{r,2}^{(1)}, w_{r,2}^{(2)}\})$  since a first round acceptance threshold of $\min\{w_{r,1}^{(1)},w_{r,1}^{(2)}\}$ is still utility-maximizing in all the previous time steps before $P$ switched to strategy $w_3$, and now a lower second round offer is utility-maximizing in the second round of all time steps where $P$ plays strategy $w_3$. The agents now agree in the second round, and $R$ prefers the second round and it is the case that $P$ prefers the first round: 

    \begin{align*}
        \min\{w_{r,1}^{(1)}, w_{r,1}^{(2)}\} - \frac{1}{D} &\le \delta (1- \min\{w_{r,2}^{(1)},w_{r,2}^{(2)}\}) \\
        1-\min\{w_{r,1}^{(1)}, w_{r,1}^{(2)}\} &\ge 1-\delta(1- \min\{w_{r,2}^{(1)},w_{r,2}^{(2)}\})-\frac{1}{D}\\
        1-\min\{w_{r,1}^{(1)}, w_{r,1}^{(2)}\} &\ge 1-\frac{1}{D}-\delta +\delta(\min\{w_{r,2}^{(1)},w_{r,2}^{(2)}\})\\
        \intertext{Since we assumed $1-\frac{1}{D} > \delta$, then we can conclude}
        1-\min\{w_{r,1}^{(1)}, w_{r,1}^{(2)}\} &> \delta(\min\{w_{r,2}^{(1)},w_{r,2}^{(2)}\})  
    \end{align*}
    Therefore, by lemma~\ref{lem:non_utility_maximizing}, there exists a time step $t_3' > t$ where $P$ switches to $(\min\{w_{r,1}^{(1)}, w_{r,1}^{(2)}\}, w_{r,2}^{(\argmax\{w_{r,1}^{(1)}, w_{r,1}^{(2)}\})})$. now we're back to strategy $w_2$ where $P$ currently prefers the first round and $R$ may prefer the first round as well and they are in NE or $R$ prefers the second round and we get the same acceptance threshold raising behavior as in the case above where $w_2$ is chosen and $R$ prefers the second round, so we can conclude this case ends in NE as well.
    \end{itemize}
    
    So, in all possible cases at $t=3$, there exists a time step $t' \ge 3$ where Algorithm~\ref{alg:modified_ftrl} eventually selects a strategy profile $(w_p^{(t')}, w_r^{(t')})$ that is in Nash Equilibrium. Finally, by lemma~\ref{lem:utility_maximizing}, the agents will stay in this strategy profile for all $t \ge t'$ and we can conclude that the algorithm has converged to a Nash Equilibrium.
\end{proof}

\begin{customex}{6}
\label{obs:n=2_disparate_outcomes}
Suppose both a proposer $f$ and responder $c$ use Algorithm 1, parameterized by $D>10, M> 2D, p=1,\delta = 0.9$ to learn in $\mathcal{G}^{(2)}$.  There exist initial conditions $w_f^{(1)}, \alpha_f, w_c^{(1)}$, and $\alpha_c$ such that the algorithms converge to a NE whose value is $\frac{D-1}{D}$.  Yet there are also initial conditions $w_{c'}^{(1)}$ and $\alpha_{c'}$ such that if the responder uses these initial conditions instead and the proposer uses $w_f^{(1)}$ and $\alpha_f$, the algorithms converge to a NE whose value is $\frac{1}{D}$.
\end{customex}
\begin{proof}[Proof sketch]
    The initial conditions we find satisfy the assumptions of Theorem~\ref{thrm:n=2_convergence}, thus the behavior of Algorithm~\ref{alg:modified_ftrl} follows as described in the theorem proof. Here, we give a high-level sketch of how the strategies change from start to convergence using lemma~\ref{lem:utility_maximizing} and lemma~\ref{lem:non_utility_maximizing} which show agents will not switch away from a utility maximizing strategy and that they will switch to a strategy that gets strictly more utility, respectively.

    First consider $f$ bargaining with $c$ when each start with:
    \begin{align*}
        w_f^{(1)} &= (0.5, w_{f,2}^{(1)} > 1-\frac{0.5}{0.9}),\\
        w_{c}^{(1)} &= (\frac{D-1}{D}, \frac{1}{D\cdot 0.9}),\\
        \text{and } \alpha_{i,1} &> w_{j,1}^{(1)} \text{ for } i \in \{f,c\}.\\ 
    \end{align*}
    
    Their $t=1$ strategies are set so that both agents prefer to make a first round deal and so we have $$(w_{f}^{(2)}, w_c^{(2)}) = \left((\frac{D-1}{D}, 1), (0.5,1)\right).$$

    At $t=3$, $c$ still prefers a first round deal, but since $0.9 > 0.5$, there exists a time $t_1$ where $f$ will switch to making a second round deal and we will have $$(w_{f}^{(t_1)}, w_c^{(t_1)}) = \left((0.48, \frac{1}{D\cdot 0.9}), (0.5,1)\right).$$
    Now, since $0.9(1-\frac{1}{D\cdot 0.9}) > 0.48$ for $D>10$, $c$ will now prefer to make a lower second round offer while $f$ prefers the current deal, so there exists a time $t_2$ where we have $$(w_{f}^{(t_2)}, w_c^{(t_2)}) = \left((0.48, \frac{1}{D\cdot 0.9}), (0.5,\frac{1}{D\cdot 0.9})\right).$$

    Now, since $0.5 > \frac{1}{D}$ for $D>10$, $f$ will prefer to make a first round deal while $c$ prefers the current deal. So, there exists a time $t_3$ where we have 
    $$(w_{f}^{(t_3)}, w_c^{(t_3)}) = \left((0.5, \frac{1}{D\cdot 0.9}), (0.5,\frac{1}{D\cdot 0.9})\right).$$
    Note that $f$ does not change their second round strategy here due to $w_c^{(1)}$. 

    Next, $c$ will prefer to get a second round deal since $0.9(1-\frac{1}{D\cdot 0.9}) > 0.5$, so, by the tie-breaking mechanism, $c$ will eventually raise their first round acceptance threshold to the next highest offer they've seen before which is $\frac{D-1}{D}$. $f$ prefers the current deal, so there exists a time $t_4$ where we have 
    $$(w_{f}^{(t_4)}, w_c^{(t_4)}) = \left((0.5, \frac{1}{D\cdot 0.9}), (\frac{D-1}{D},\frac{1}{D\cdot 0.9})\right).$$

    Finally, since $c$ is back to playing their first round strategy, $f$ will prefer a first round deal and they will switch back to their highest offer (and accept only 1 in the second round by the tie-breaking mechanism) and there exists a time $t_5$ where the agents finally end in NE:
    $$(w_{f}^{(t_5)}, w_c^{(t_5)}) = \left((\frac{D-1}{D}, 1), (\frac{D-1}{D},\frac{1}{D\cdot 0.9})\right).$$

    Next, consider $f$ bargaining with $c$ when each start with:
    \begin{align*}
        w_f^{(1)} &= (0.5, w_{f,2}^{(1)} > 1-\frac{0.5}{0.9}),\\
        w_{c}^{(1)} &= (\frac{1}{D}, 1),\\
        \text{and } \alpha_{i,1} &> w_{j,1}^{(1)} \text{ for } i \in \{f,c\}.\\ 
    \end{align*}. 
    
    Again, their $t=1$ strategies are set so that both agents prefer to make a first round deal and so we have $$(w_{f}^{(2)}, w_{c_2}^{(2)}) = \left((\frac{1}{D}, 1), (0.5,1)\right).$$

    Now, $f$ prefers the second round deal they are currently getting since $0.9 > 0.5$, but $c$ would prefer to get something rather than nothing, so there exists a time $t_1$ where agents switch to the following profile in NE:
    $$(w_{f}^{(t_1)}, w_{c_2}^{(t_1)}) = \left((\frac{1}{D}, 1), (\frac{1}{D},1)\right).$$
\end{proof}

Even though $f$ had an initial offer of $0.5$ to both agents, the initial acceptance threshold of each agent ends up determining the final NE convergence value, though for different reasons. For $c_1$, acceptance of their second round offer is preferable over $f$'s offer of $w_{f,1}^{(1)}$, so $c_1$ eventually raises their initial acceptance threshold back to their initial threshold of $\frac{D-1}{D}$ to force the second round deal. $f$ eventually matches the first round acceptance threshold and then responds with a threat of their own to take everything in the second round as a result of the tie-breaking mechanism. On the other hand, for $c_2$, $f$ matches their low acceptance threshold early on and offers nothing in the second round again due to the tie-breaking mechanism and $c_2$ must accept a deal with their initial low acceptance threshold of $\frac{1}{D}$ to avoid getting nothing. 

\newpage
\subsection{Additional Simulation Results}

This section gives additional empirical convergence results from simulating Algorithm~\ref{alg:modified_ftrl} using CVXPY~\citep{diamond2016cvxpy} over a range of initial conditions. The parameters we use are specified beneath each figure. Since we use $p=1$ for the regularizer, we ensure only pure strategies are chosen at each time step by rounding to 5 decimal places and using the tie-breaking mechanism described above.  

\begin{figure*}[t!h]
    \centering
    \begin{subfigure}[t]{0.4\textwidth}
    
        \centering
        \includegraphics[height=2.5in]{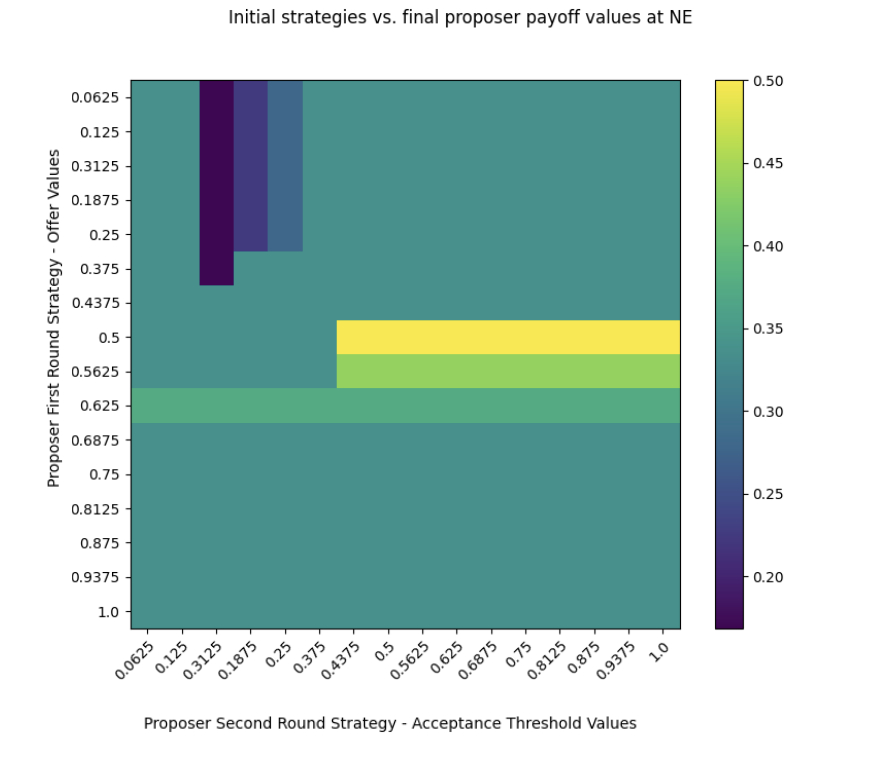}
        \caption{Simulation results for NE outcomes of agents learning strategies in $\mathcal{G}^{(2)}$. The setting is $T=100, M=40, D=16, p=1, \alpha_P = (0.125, 0.375), \alpha_R = (0.375, 0.875), w_R^{(1)} = (0.5, 0.375)$. The initial strategy of the proposer varies from $\{\frac{1}{D},\ldots, 1\}$ in both strategy dimensions.}
        \label{fig:fixed_responder1}
    \end{subfigure}
    \hfill
    \begin{subfigure}[t]{0.4\textwidth}
    
        \centering
        \includegraphics[height=2.5in]{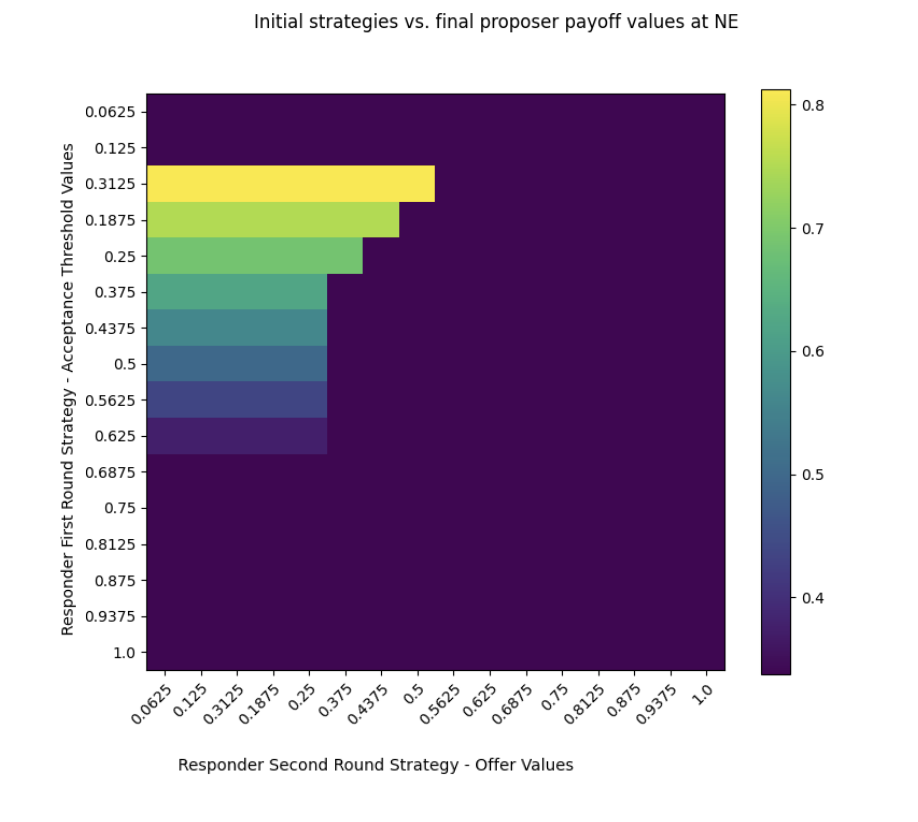}
        \caption{Simulation results for NE outcomes of agents learning strategies in $\mathcal{G}^{(2)}$. The setting is $T=100, M=40, D=16, p=1, \alpha_P = (0.125, 0.375), \alpha_R = (0.375, 0.875), w_P^{(1)} = (0.375, 0.125)$. The initial strategy of the responder varies from $\{\frac{1}{D},\ldots, 1\}$ in both strategy dimensions.}
        \label{fig:fixed_proposer1}
    \end{subfigure}
    \hfill
    \begin{subfigure}[t]{0.4\textwidth}
    
        \centering
        \includegraphics[height=2.5in]{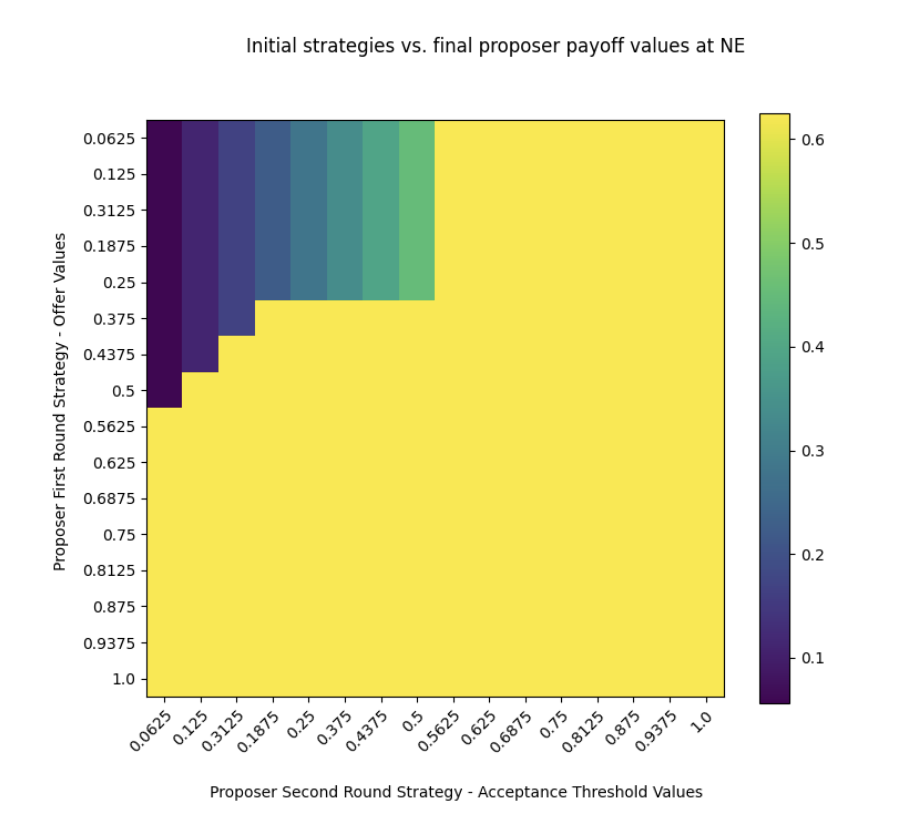}
        \caption{Simulation results for NE outcomes of agents learning strategies in $\mathcal{G}^{(2)}$. The setting is $T=100, M=40, D=16, p=1, \alpha_P = (0.125, 0.375), \alpha_R = (0.375, 0.875), w_R^{(1)} = (0.375, 0.125)$. The initial strategy of the proposer varies from $\{\frac{1}{D},\ldots, 1\}$ in both strategy dimensions.}
        \label{fig:fixed_responder2}
    \end{subfigure}
    \hfill
    \begin{subfigure}[t]{0.4\textwidth}
    
        \centering
        \includegraphics[height=2.5in]{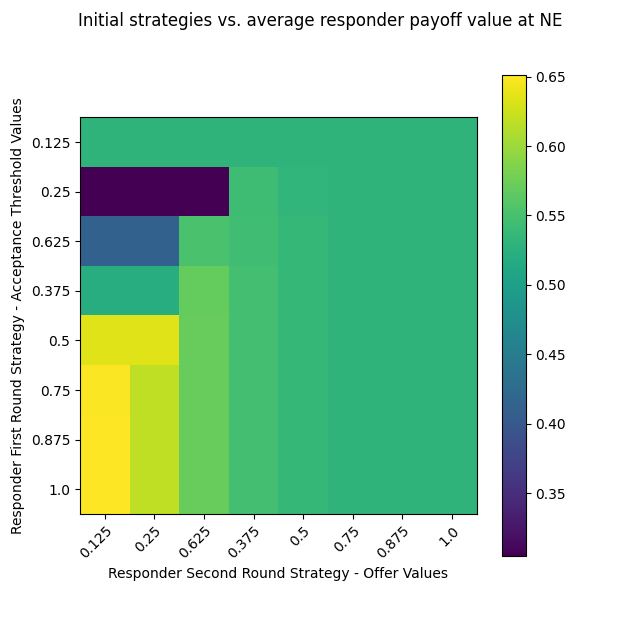}
        \caption{Simulation results for NE outcomes of agents learning strategies in $\mathcal{G}^{(2)}$. The setting is $T=300, M=40, D=16, p=1, \alpha_P = (0.125, 0.375), \alpha_R = (0.375, 0.875)$. The initial strategy of the responder varies from $\{\frac{1}{D},\frac{3}{D}\ldots, \frac{D-1}{D}\}$ in both strategy dimensions. The color of each cell represents the average payoff to the responder playing that initial strategy over the initial strategies the proposer plays from the same set.}
        \label{fig:avg_responder}
    \end{subfigure}
\end{figure*}

We first chose a subset of strategy values from $\{0,\frac{1}{D},\ldots,1\}$ and fixed reference points. Then, we ran the algorithm for each possible pairing of initial strategies of the proposer and the responder from this subset (Figure~\ref{fig:avg_responder}) or we fixed one of the proposer or responder and varied the initial conditions of the other (Figures~\ref{fig:fixed_proposer1}~\ref{fig:fixed_responder1}~\ref{fig:fixed_responder2}). Empirically, we observe that the algorithm converges to NE within $T=300$ time steps for all of these initial conditions, and the color of each cell either corresponds to an average payoff or payoff given a fixed proposer or responder strategy (see captions).

\end{document}